\documentclass[journal,twocolumn,print]{IEEEtran}
\usepackage{cite}
\usepackage{scalerel}
\usepackage{amsmath,amssymb,amsfonts}
\usepackage{algorithmic}
\usepackage{graphicx}
\usepackage{textcomp}
\usepackage[dvipsnames]{xcolor}
\DeclareGraphicsExtensions{.pdf}
\usepackage{tabularx, booktabs}
\usepackage[switch]{lineno}
\renewenvironment{thebibliography}[1]{
  \begin{oldthebibliography}{#1}
    \setlength{\itemsep}{0.40em}
    \setlength{\parskip}{-0.45em}
}
{
  \end{oldthebibliography}
}

\def\BibTeX{{\rm B\kern-.05em{\sc i\kern-.025em b}\kern-.08em
T\kern-.1667em\lower.7ex\hbox{E}\kern-.125emX}}

\makeatletter
\renewcommand{\@maketitle}{%
    \newpage\null\vskip1em%
    \begin{center}%
        \let\footnote\thanks{\Huge\@title\par}%
        \vskip0.5em{\normalsize\lineskip0.10em\begin{tabular}[t]{c}\@author\end{tabular} \par }%
        \vspace{-0.45in}
    \end{center}%
    \par}
\makeatother

\begin{document}
\title{LungViT: Ensembling Cascade of Texture Sensitive Hierarchical Vision Transformers for Cross-Volume Chest CT Image-to-Image Translation}
\author{Muhammad F. A. Chaudhary, Sarah E. Gerard, Gary E. Christensen, Christopher B. Cooper, \\ Joyce D. Schroeder, Eric A. Hoffman, and Joseph M. Reinhardt 
\thanks{Muhammad F. A. Chaudhary, Sarah E. Gerard, and Joseph M. Reinhardt are with The Roy J. Carver Department
of Biomedical Engineering, The University of Iowa, Iowa City, IA, 52242, USA. e-mail: joe-reinhardt@uiowa.edu}
\thanks{Gary E. Christensen is with the Department of Electrical and Computer Engineering, The University of Iowa, Iowa City, IA, USA, 52242.}
\thanks{Christopher B. Cooper is with the Department of Physiology, David Geffen School of Medicine, University of California, Los Angeles, CA, USA, 90095.}
\thanks{Joyce D. Schroeder is with the Department of Radiology and Imaging Sciences, The University of Utah, Salt Lake City, UT, USA, 84132.}
\thanks{Eric A. Hoffman is with the Department of Radiology, The University of Iowa, Iowa City, IA, 52242, USA.}}

\maketitle

\begin{abstract}
Chest computed tomography (CT) at inspiration is often complemented by an expiratory CT to identify peripheral airways disease. Additionally, co-registered inspiratory-expiratory volumes can be used to derive various markers of lung function. Expiratory CT scans, however, may not be acquired due to dose or scan time considerations or may be inadequate due to motion or insufficient exhale; leading to a missed opportunity to evaluate underlying small airways disease. Here, we propose LungViT -- a generative adversarial learning approach using hierarchical vision transformers for translating inspiratory CT intensities to corresponding expiratory CT intensities. LungViT addresses several limitations of the traditional generative models including slicewise discontinuities, limited size of generated volumes, and their inability to model texture transfer at volumetric level. We propose a shifted-window hierarchical vision transformer architecture with squeeze-and-excitation decoder blocks for modeling dependencies between features. We also propose a multiview texture similarity distance metric for texture and style transfer in 3D. To incorporate global information into the training process and refine the output of our model, we use ensemble cascading. LungViT is able to generate large 3D volumes of size 320 $\times$ 320 $\times$ 320. We train and validate our model using a diverse cohort of 1500 subjects with varying disease severity. To assess model generalizability beyond the development set biases, we evaluate our model on an out-of-distribution external validation set of 200 subjects. Clinical validation on internal and external testing sets shows that synthetic volumes could be reliably adopted for deriving clinical endpoints of chronic obstructive pulmonary disease.  
\end{abstract}

\begin{IEEEkeywords}
Generative adversarial networks (GANs), medical image synthesis, style transfer, texture, lungs, vision transformers
\end{IEEEkeywords}
\vspace{-0.10in}

\section{Introduction}
\label{sec:introduction}
\IEEEPARstart{C}{hest} computed tomography (CT) is the imaging modality most commonly used to diagnose lung disease~\cite{chaudhary2023predicting,bodduluri2018recent,amudala2023radiomics}. Advancements in CT hardware have enabled acquisition of large volumes that capture anatomical details at high resolution~\cite{hoffman2022origins}. Computed tomography, acquired at multiple lung volumes, facilitates clinical decision making by offering complementary information about local anatomical structure and function. Deformable image registration (DIR), typically of the end-inspiratory and end-expiratory CT volumes, is used to derive various surrogates for regional lung function~\cite{reinhardt2008registration,galban2012computed,kirby2017novel}. Such multivolume measures have improved our understanding of pulmonary abnormalities such as lung cancer and chronic obstructive pulmonary disease (COPD)~\cite{reinhardt2008registration,galban2009parametric,galban2012computed,kirby2017novel,bodduluri2017biomechanical}. For instance, parametric response mapping (PRM) and disease probability maps (DPM) have been used to identify local patterns of emphysema and functional small airways disease (fSAD) from co-registered, inspiratory-expiratory chest CT volumes~\cite{galban2012computed,kirby2017novel}. PRM and DPM have since gathered widespread clinical attention as they have been used to detect and characterize COPD and asthma~\cite{trivedi2022quantitative}. Bodduluri \textit{et al}. showed that registration-derived local tissue expansion of the lung was associated with respiratory function, emphysema, and patient quality of life~\cite{bodduluri2017biomechanical}. 

While multiple volumes improve clinical assessment, acquiring them takes additional time and results in increased radiation dose and overall cost. Each high-resolution chest CT volume acquired at inspiration can expose a patient to several milli-Gray of ionizing radiation. An additional chest CT volume at expiration entails at least half of the radiation dosage of an inspiratory scan~\cite{sieren2016spiromics,newell2013development}. The radiation exposure may increase even further in cases where subjects are not able to achieve reliable end-inspiratory or end-expiratory chest CT volumes and may need a repeat scan. These cases include advanced-stage COPD and lung cancer patients with increased disease burden. Expiration imaging is not routinely ordered by clinicians, who may not be aware of the value of inspiration/expiration paired data. Several large multicenter studies acquired CT scans at different lung volumes making it difficult to characterize subjects across different cohorts. For instance, the multi-ethnic study of atherosclerosis (MESA) acquired CT scans at total lung capacity (TLC) only~\cite{bild2002multi} and the Genetic Epidemiology of COPD (COPDGene) study acquired CT scans at TLC and functional residual capacity (FRC)~\cite{regan2011genetic}. These problems limit the widespread utility and retrospective evaluation of multivolume approaches for analyzing lung disease and characterizing patterns across cohorts. Cross-volume image synthesis could serve as a useful tool towards mitigating this challenge. Having the ability to synthesize expiratory CT images could allow for retrospective evaluation of lung function when legacy scans consist only of an inspiratory data set or the expiratory images were corrupted by an inadequate expiratory effort or motion, which is often the case.

In recent years, generative models have experienced a surge in popularity for various medical image analysis tasks including, image synthesis and translation~\cite{wolterink2017deep,nie2018medical,yu2019ea,yang2020unsupervised,yu2020sample,zhou2020hi,dar2019image,emami2018generating,uzunova2020memory}, reconstruction~\cite{luo20213d,zhang2021transct,korkmaz2022unsupervised}, super-resolution~\cite{you2019ct,you2022fine,lyu2020mri}, denoising~\cite{yang2018low,bera2021noise,zhang2021transct}, and data augmentation~\cite{shin2018medical,frid2018gan}. Earlier works in this area used 2D slice-based generative adversarial networks (GANs) for low-dose CT denoising~\cite{yang2018low}, motion correction~\cite{kustner2019retrospective}, and converting magnetic resonance slices to synthetic CT slices~\cite{wolterink2017deep}. However, 2D GANs typically suffer from slicewise discontinuities which may not suitable for radiological purposes. Volumetric GANs address this issue by processing 3D volumes instead of 2D images. For instance, Nie \textit{et al}. used a 3D GAN with auto-context refinement for CT patch synthesis, achieving consistency across all spatial dimensions~\cite{nie2018medical}. Edge-aware GANs used 3D edge priors for multimodal magnetic resonance image synthesis~\cite{yu2019ea}. Volumetric GANs require large GPU memory for training and most approaches either train their models on patches or downsampled volumes. Training 3D generators on patches can lead to undesirable results due to lack of global context. This is further aggravated by the inherently local nature of fully convolutional generators. Due to these reasons most 3D generative models were restricted to limited volume sizes.

Volumetric approaches often rely on carefully designed image priors or loss functions to address the lack of context in 3D patch-based networks~\cite{nie2017medical,nie2018medical,yu2019ea}. Although such strategies improve model performance, they fail to directly address network locality. Vision transformers (ViTs) have recently emerged as an alternative to purely convolutional architectures since they are able to learn contextual representations through self-attention~\cite{dosovitskiy2020image}. Recent examples like VTGAN~\cite{kamran2021vtgan} for retinal image synthesis, ResViT for unified multimodal medical image synthesis~\cite{dalmaz2022resvit}, and PTNet for infant brain magnetic resonance image synthesis~\cite{zhang2022ptnet3d} demonstrate improved performance of ViT-based architectures over fully convolutional methods. Most of these approaches still resort to slice-based training due to significantly increased GPU memory demands of ViTs. Another important limitation presented by most of the GAN studies is a lack of model validation against clinical end-points and on large out-of-distribution external validation sets -- a component which was completely missing in all the studies discussed above. For our task of cross-volume CT synthesis, it is pertinent to generate high-resolution 3D volumes and validate them against clinical markers of disease for broader clinical applicability. 

We propose LungViT, a texture sensitive, multiscale vision transformer-based approach for predicting inspiratory-to-expiratory intensity change in chest CT. We hypothesize that a CT image at inspiration contains sufficient structural information to predict the associated aeration change on an expiratory CT image. We propose a hybrid generator architecture called SwinSEER, which uses shifted-window (Swin) transformers within the the encoder, and squeeze-and-excitation (SE) blocks with convolutions within the decoder. Unlike most GAN approaches, we optimize a multiview texture- and structure-sensitive image quality assessment metric for improving texture synthesis and increase robustness to structural distortions. We show that LungViT is able to capture subtle tissue textures and generate high-resolution CT volumes up to a size of 320 $\times$ 320 $\times$ 320 -- significantly larger than the volumes generated by most of the recent 3D GAN approaches. The major contributions of our work can be summarized as follows. 

\begin{itemize}
\item We believe this is the first study that investigates the role of generative models for cross-volume computed tomography.
\item We propose a hybrid 3D generator called SwinSEER, based on convolutional and transformer blocks, for synthesizing large 3D medical image volumes. 
\item We propose a multiview extension of the deep image structure and texture similarity (DISTS) metric~\cite{ding2020image} for modeling tissue texture and style transfer across volumes.
\item We also propose ensemble cascading to better capture global context for volumetric image synthesis; and compare the performance of our method with various state-of-the-art 2D and 3D image-to-image translation methods including both convolutional and transformer architectures.
\item Unlike most generative modeling studies, we demonstrate that synthetic images, derived using LungViT, can be used to reliably compute several clinical end-points of lung function.
\item We assess model generalizability on a large out-of-distribution external validation cohort where CT scans were acquired using a totally different protocol.
\end{itemize}

\begin{figure*}[!t]
\centering
\includegraphics[width=1.05\textwidth]{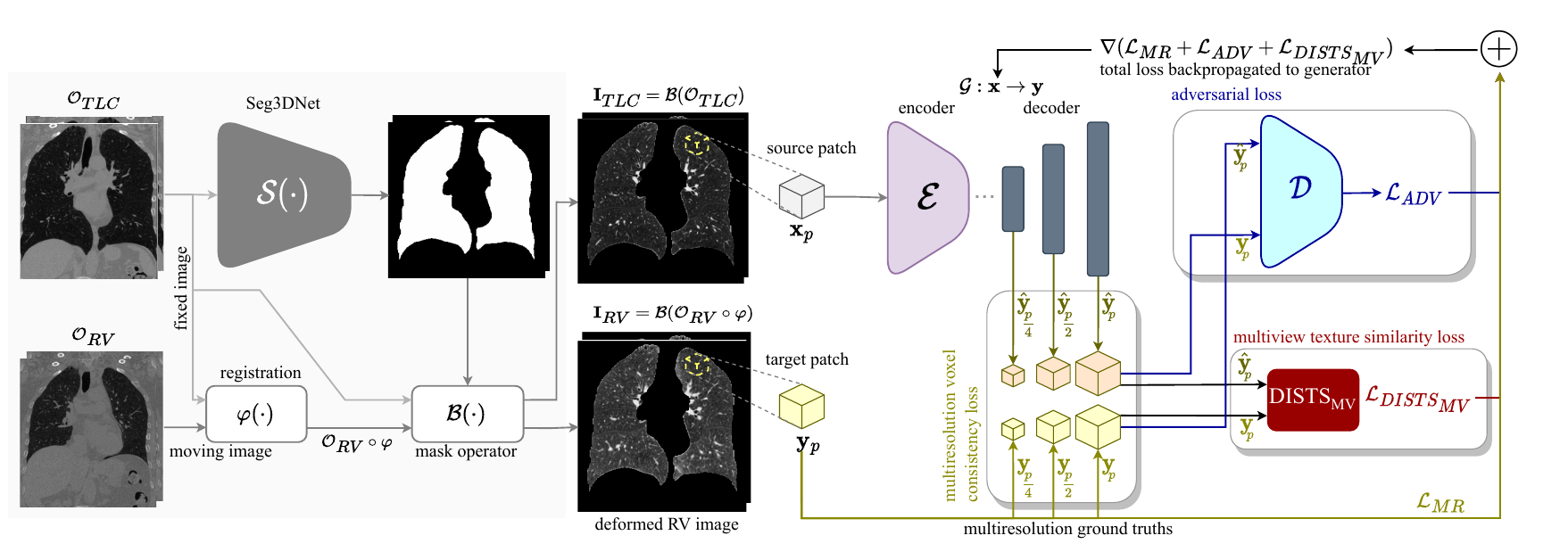}
\vspace{-0.15in}
\caption{Overview of the LungViT framework. We registered CT volumes at inspiration $\mathcal{O}_{\mathrm{TLC}}$ (fixed image) and expiration $\mathcal{O}_{\mathrm{RV}}$ (moving image) using a deformable image registration (DIR) method. The lung segmentation mask for the original TLC image $\mathcal{O}_{\mathrm{TLC}}$ was obtained using Seg3DNet~\cite{gerard2020multi}. The operator from TLC mask $\mathcal{B}(\cdot)$ was used to generate two masked volumes -- $\boldsymbol{I}_{\mathrm{TLC}} = \mathcal{B}(\mathcal{O}_{\mathrm{TLC}})$ and $\boldsymbol{I}_{\mathrm{RV}} = \mathcal{B}(\mathcal{O}_{\mathrm{RV}} \circ \varphi)$ (deformed to the TLC coordinate system using the transformation $\varphi$ estimated by DIR). A generator $\mathcal{G}$ and discriminator $\mathcal{D}$ were then trained on 3D patches $\mathbf{x}_{p}$ and $\mathbf{y}_{p}$, of size $\mathrm{p} \times \mathrm{p} \times \mathrm{p}$, extracted from the masked $\boldsymbol{I}_{\mathrm{TLC}}$ and $\boldsymbol{I}_{\mathrm{RV}}$ volumes (both in TLC space) respectively. Notice, in addition to the adversarial loss $\mathcal{L}_{\mathrm{ADV}}$, gradients from the multiview texture similarity module $\nabla \mathcal{L}_{\mathrm{DISTS_{MV}}}$ and multiresolution voxel consistency loss $\nabla \mathcal{L}_{\mathrm{MR}}$ were also back-propagated for learning the generator $\mathcal{G}$. `$\mathcal{M}_1 \dots \mathcal{M}_4$' indicate 3D convolutions with unit filter size (1 $\times$ 1 $\times$ 1).}
\label{overview}
\end{figure*}

\section{Related Work}
Generative adversarial network architectures were initially based on fully convolutional generators and discriminators. Vision transformer-based architectures were recently demonstrated to perform better at medical image synthesis tasks. In this section, we review both the paradigms and point out some of their pertinent limitations.

\subsection{Convolutional Generative Models}
Fully Convolutional GANs have been shown to perform remarkably well for various medical image processing tasks such as denoising~\cite{yang2018low,bera2021noise}, super-resolution~\cite{you2019ct,you2022fine,lyu2020mri}, data augmentation~\cite{shin2018medical,frid2018gan}, and image-to-image translation~\cite{wolterink2017deep,nie2018medical,yu2019ea,yang2020unsupervised,yu2020sample,zhou2020hi,dar2019image,emami2018generating,uzunova2020memory}. The GAN frameworks can be differentiated by the number of spatial dimensions they attempts to model, i.e., planar (2D) vs. volumetric (3D) approaches. 

Most of the preliminary work done towards medical image-to-image translation was performed by 2D GAN frameworks that operated on 2D slices instead of 3D volumes. For instance, Yang \textit{et al}. proposed to jointly minimize adversarial and perceptual loss functions for denoising low-dose CT slices from paired high-dose samples~\cite{yang2018low}. Similarly, Armanious \textit{et al}. proposed MedGAN, which is a cascade of 2D UNet blocks, trained by minimizing adversarial and perceptual loss functions~\cite{armanious2020medgan}. The MedGAN architecture was also used for motion correction of magnetic resonance images~\cite{kustner2019retrospective}. In~\cite{emami2018generating}, an image-conditional GAN was trained in a supervised fashion to convert T$_{\mathrm{1}}$-weighted magnetic resonance slices to synthetic CT slices. Zhou \textit{et al}. developed a feature fusion approach for cross-modal magnetic resonance image synthesis~\cite{zhou2020hi}. They trained three different network backbones using paired 2D slices to fuse representations from different modalities. The multiple-backbone architecture demanded larger memory and the framework was hence constrained to 2D slices~\cite{zhou2020hi}. Lyu \textit{et al}. proposed an ensemble learning approach for magnetic resonance image super-resolution. They improved model performance by using upsampled image priors for training their GAN~\cite{lyu2020mri}. Küstner \textit{et al}. trained a cascade of motion-compensated 2D UNet generators for cardiac angiogram super-resolution~\cite{kustner2021deep}. A discriminator rejection sampling-based 2D GAN was used to estimate local lung tissue mechanics from a single inspiratory chest CT~\cite{chaudhary2022single}. Wolterlink \textit{et al}. used a CycleGAN~\cite{zhu2017unpaired} for unpaired magnetic resonance to CT slice translation. This modeling framework relaxed the assumption of requiring paired training data and utilized two generative models to ensure cycle-consistency between domains~\cite{wolterink2017deep}. The lack of supervised training in unpaired approaches often leads to structural inconsistencies between real and synthetic images. Recently, Yang \textit{et al}. tried to address this by exploiting common structural attributes between two modalities~\cite{yang2020unsupervised}. All of these approaches used 2D slices for training which appear discontinuous when stacked together into a volume. The slice-based methods are unable to explicitly learn large-scale 3D texture patterns, which strictly reduces the clinical utility of the synthetic images. The unpaired approaches that rely on multiple generative models incur large GPU memory for training and are harder to scale at volumetric levels.

Volumetric (3D) generative frameworks address several limitations presented by the slice-based (2D) methods~\cite{nie2018medical,shin2018medical,yu2019ea,yu2020sample,uzunova2020memory}. Nie \textit{et al}. trained a 3D GAN on magnetic resonance image patches to generate corresponding CT patches. They further used auto-context~\cite{tu2009auto} to iteratively refine CT patch synthesis by conditioning each successive GAN on outputs from the previous model~\cite{nie2018medical}. The networks operated on small image patches (of size 32 $\times$ 32 $\times$ 32) and the overall volume size generated using the trained model was limited to 153 $\times$ 193 $\times$ 50. The framework additionally proposed to minimize the gradient difference loss to encourage structural consistency between real and synthetic samples~\cite{nie2018medical}. Unlike results from 2D GANs, the synthetic CT volumes were consistent across all three spatial dimensions~\cite{nie2018medical}. Shin \textit{et al}. also proposed a 3D GAN for data augmentation but had to downsample their image volumes to 128 $\times$ 128 $\times$ 54 due to limited GPU memory available for training~\cite{shin2018medical}. Recently, another patch-based 3D approach, the edge-aware GANs, leveraged edge detection to better capture structural features for multi-modality MR image synthesis~\cite{yu2019ea}. The trained model was demonstrated to generate an overall volume of size 240 $\times$ 240 $\times$ 155. A multi-resolution cascade of GANs was also proposed to alleviate the memory requirements for synthesizing high-resolution 3D volumes~\cite{uzunova2020memory}. Similar to Nie \textit{et al}., this approach used contextual information from a lower resolution model to train a cascade of patch-based higher resolution models. Volumetric generative models have been successfully applied for medical image super-resolution as well. Chen \textit{et al}. proposed a memory efficient, multi-level densely connected super-resolution generator. Each dense block was designed to have dense connections and compressors, which helped improve the overall memory requirement of the proposed framework~\cite{chen2020mri}. Similarly, Pham \textit{et al}. investigated the utility of multiscale 3D CNNs for mono-modal and multi-modal medical image super-resolution~\cite{pham2019multiscale}. A volumetric (3D) counterpart of the well-known 2D image-to-image translation method, Pix2Pix~\cite{isola2017image}, was used for correcting rigid-body motion artifacts in dynamic magnetic resonance images\cite{johnson2019conditional}. A 3D convolutional generator, Reg3DNet+, was recently used for estimating lung tissue mechanics from inspiratory and expiratory CT volumes~\cite{gerard2023direct}. 

\subsection{Transformer-based GANs}
Although fully convolutional generator architectures have been applied to various medical image analysis problems, they are inherently local and are unable to capture long-range dependencies between pixels of an image. This limitation was first addressed, in part, by using self-attention modules in a GAN (SAGAN) generator~\cite{zhang2019self}. Li \textit{et al}. used a convolutional generator with pseudo-3D self-attention (SACNN) for low-dose CT denoising~\cite{li2020sacnn}. A major shortcoming of the self-attention module, however, is that it requires significantly larger GPU memory for modeling the long-range dependencies in an image, making it increasingly difficult to handle larger image sizes and 3D volumes. This limitation was recently addressed using a vision transformer, a non-convolutional generator architecture that processes low-dimensional patch embeddings for learning long-range dependencies between them~\cite{dosovitskiy2020image}. Due to their ability to learn improved contextual representations, transformer-based methods have gained recent attention for various medical imaging tasks including segmentation~\cite{chen2021transunet,xie2021cotr,hatamizadeh2022unetr}, synthesis~\cite{kamran2021vtgan,shin2020ganbert,zhang2022ptnet3d,li2023multi}, registration~\cite{chen2022transmorph,shi2022xmorpher}, and reconstruction~\cite{luo20213d,zhang2021transct,korkmaz2022unsupervised}. 

A coarse-to-fine, pure transformer GAN was proposed for 2D retinal image synthesis and disease diagnosis~\cite{kamran2021vtgan}.
A hybrid convolutional and transformer-based architecture, called ResViT, was developed for unified multimodal magnetic resonance image synthesis~\cite{dalmaz2022resvit}. The proposed generator used residual transformer blocks in the bottleneck layers to learn global and local image features. The method was trained on 2D slices due to GPU memory constraints~\cite{dalmaz2022resvit}. More recently, a multiscale pure transformer architecture, MTNet, was proposed for multimodal MR image synthesis~\cite{li2023multi}. This framework also suffered from large GPU memory requirements and was trained on 2D slices. A 3D hybrid model called GANBERT was used for synthesizing positron emission tomography images from the corresponding magnetic resonance images~\cite{shin2020ganbert}. Like other 3D convolutional approaches, the generated volume size was limited to 93 $\times$ 76 $\times$ 76. Zhang \textit{et al}. proposed a pure transformer-based pyramical generative architecture for magnetic resonance image synthesis in infant brains~\cite{zhang2022ptnet3d}. Although the framework processed 3D volumes, it demanded significant GPU memory and was trained on small patches of size 64 $\times$ 64 $\times$ 64. 

The current state of hybrid and pure transformer architectures clearly points towards a major limitation -- limited size of generated volumes due to GPU memory constraints. This is further aggravated by the increasingly large 3D chest CT volumes acquired for diagnosing lung disease. Another pertinent limitation of almost all the models discussed above is a lack of clinical validation by downstream analyses and external validation on out-of-distribution datasets.

\subsection{Neural Style and Texture Transfer}
Neural style transfer is an underexplored paradigm in medical image synthesis. This can be attributed to a lack of 3D pre-trained networks for generating style embeddings and also to the absence of stylization exemplars for medical images. Most of the literature for medical image-to-image translation used learned perceptual image patch similarity (LPIPS)~\cite{zhang2018unreasonable} for improving the perceptual quality of the generated images~\cite{yang2018low,armanious2020medgan}. Recently, Liu \textit{et al}. used stylization codes for magnetic resonance image harmonization~\cite{liu2021style}. For our problem of cross-volume CT synthesis it is pertinent to model lung tissue style and texture transfer during image synthesis.

\section{Background}
Generative adversarial networks (or GANs) implicitly model the data distribution by learning a mapping $\mathcal{G}: \mathcal{Z} \to \mathcal{X}$ from a random noise vector $\mathbf{z} \in \mathcal{Z}$ to a given image $\mathbf{x} \in \mathcal{X}$, where $\mathbf{x}$ is sampled from the data distribution~\cite{goodfellow2014generative}. The GAN framework consists of two different networks that are trained to compete against each other until they reach Nash's equilibrium~\cite{goodfellow2014generative}. This game-theoretic training approach has been successfully used to generate perceptually realistic images for various tasks. Isola \textit{et al}. extended the original GAN framework to the image conditional GAN (cGAN) that is able to learn a generator $\mathcal{G}: \mathcal{X} \to \mathcal{Y}$, from an image $\mathbf{x}$ (in domain $\mathcal{X}$) to another image $\mathbf{y}$ (in domain $\mathcal{Y}$)~\cite{isola2017image}. The synthetic samples $\mathbf{\hat{y}} = \mathcal{G}(\mathbf{x})$, generated by $\mathcal{G}$ are then evaluated by a second model called the discriminator $\mathcal{D}$, which is simultaneously trained to learn if the samples are real or fake. Typically, $\mathcal{D}$ is a binary classifier that outputs the probability of a sample being derived from the target (or data) distribution. While $\mathcal{D}$ is optimized to maximize this probability, the generator $\mathcal{G}$ tries to minimize it, which creates a dynamic modeled by the minimax game stated as:
\begin{equation}
\label{cgan}
\mathcal{L}_{\mathrm{cGAN}} = \mathbb{E}_{\mathbf{x,y}}[\mathrm{log}\:\mathcal{D}(\mathbf{x,y})] \: + \mathbb{E}_{\mathbf{x}}[\mathrm{log}\:(1 - \mathcal{D}(\mathbf{x}, \mathcal{G}(\mathbf{x})))],  
\end{equation}
where $\mathcal{G}$ is trained to minimize (\ref{cgan}) such that $\{\mathcal{G}^{\star}\} = \operatorname{min}_{\mathcal{G}} \operatorname{max}_{\mathcal{D}}\mathcal{L}_{\mathrm{cGAN}} $. Isola \textit{et al}. suggested to additionally minimize the pixelwise distance ($\ell_1$ or $\ell_2$) between the real $\mathbf{y}$ and generated samples $\mathbf{\hat{y}}$~\cite{isola2017image}. The minimax objective in (\ref{cgan}) can be extended to express the overall generator $\mathcal{G}$ loss function as:
\begin{equation}
    \label{cgan_G}
    \mathcal{L}_{\mathcal{G}}(\mathbf{x},\: \mathbf{y}) = \underbrace{\mathcal{L}_{\mathrm{BCE}}(\mathcal{D}(\mathbf{x},\: \mathcal{G}(\mathbf{x})), 1)}_\text{cross entropy adversarial loss~} + \underbrace{\lambda \mathbb{E}[\lVert \mathbf{y} - \mathcal{G}(\mathbf{x}) \rVert_{1}] }_\text{$\ell_1$ -- distance~},
\end{equation}
where the first term in (\ref{cgan_G}) refers to binary cross-entropy (BCE) loss minimized by $\mathcal{G}$, and the second term constitutes the $\ell_1$ -- distance between the real $\mathbf{y}$ and generated samples $\mathbf{\hat{y}}$. The aforementioned framework is called Pix2Pix and has been shown to be very effective for various 2D image-to-image translation tasks~\cite{isola2017image}.

\begin{figure*}[!t]
\centering
\includegraphics[width=1.02\textwidth]{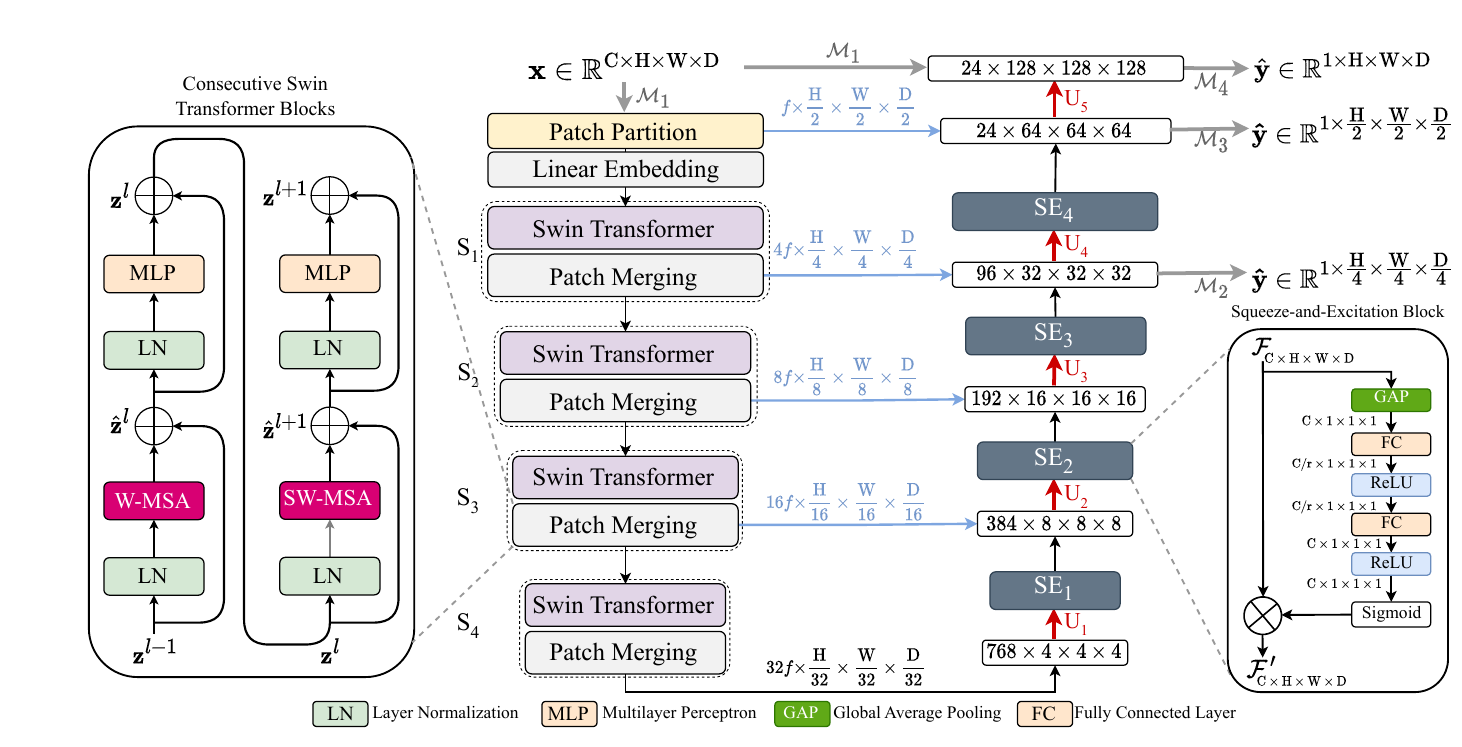}
\vspace{-0.35in}
\caption{The proposed LungViT generator. The encoder consists of four hierarchical feature processing Swin Transformer blocks, $\mathrm{\{S_1, S_2, S_3, S_4\}}$, that learn image representations at multiple resolutions. The feature dimensions $f$, initially set to 24, increase down the bottleneck up to 768, while the spatial dimensions decrease to $4 \times 4 \times 4$. All stages use 2 Swin Transformer blocks, except $\mathrm{S}_2$, which uses 4 blocks. We have also shown an expanded structure of the Swin Transformer block besides the LungViT encoder. The decoder receives input from the bottleneck and is also connected to different stages of the encoder via skip connections. The decoder backbone consists of alternation convolutional and SE blocks. The decoder outputs the generated image at three different resolutions to be used by the multiresolution loss. For the convolutional blocks in the decoder $\{\mathrm{U}_1 \dots \mathrm{U}_5\}$, we use instance normalization. The internal architecture of the SE block is also shown to the right of LungViT decoder.}
\label{gen}
\end{figure*}

\section{LungViT}
\subsection{Overview}
We introduce a neural style and texture transfer approach for translating inspiratory CT image intensities to corresponding expiratory CT image intensities (see Fig. \ref{overview}). The training framework was preceded by registration and segmentation of the inspiratory CT image acquired at total lung capacity (TLC), denoted as $\mathcal{O}_{\mathrm{TLC}}$, and expiratory CT image acquired at residual volume (RV), denoted as $\mathcal{O}_{\mathrm{RV}}$ (see Fig. \ref{overview}). We used a tissue intensity and fissure structure constrained deformable image registration (DIR) method to match $\mathcal{O}_{\mathrm{RV}}$ (treated as the moving image) to $\mathcal{O}_{\mathrm{TLC}}$ (treated as the fixed image)~\cite{yin2009mass}. Lung segmentation mask of the original TLC image $\mathcal{O}_{\mathrm{TLC}}$ was obtained using a cascaded, multiscale CNN framework that uses the Seg3DNet architecture~\cite{gerard2020multi}. As shown in Fig. \ref{overview}, the operator from TLC image mask was used to generate two masked volumes -- $\boldsymbol{I}_{\mathrm{TLC}}$ and $\boldsymbol{I}_{\mathrm{RV}}$ where $\boldsymbol{I}_{\mathrm{RV}}$ corresponded to the RV image deformed to the TLC image space. 

We trained a volumetric image conditional least squares GAN on paired image patches ($\mathbf{x}_p$ and $\mathbf{y}_p$) extracted from the co-registered, masked volumes, $\boldsymbol{I}_{\mathrm{TLC}}$ and $\boldsymbol{I}_{\mathrm{RV}}$, respectively. The generator architecture, called SwinSEER, consisted of a multiresolution shifted-window vision transformer-based encoder for modeling long-range dependencies between distant image regions. Its decoder consisted of transposed convolutions and squeeze-and-excitation blocks  for adaptive recalibration of feature maps through channel attention. To harness the advantages of one-to-one correspondence between source (TLC patches or $\mathbf{x}_p$) and target domain images (RV patches or $\mathbf{y}_p$), we trained our models in a paired image-to-image translation setting. This allowed us to develop a multiresolution voxel consistency loss ($\mathcal{L}_{\mathrm{MR}}$) that encouraged high-resolution image synthesis. In addition to the $\mathcal{L}_{\mathrm{MR}}$ and the adversarial loss $\mathcal{L}_{\mathrm{ADV}}$ from discriminator $\mathcal{D}$, we developed a multiview DISTS loss $\mathcal{L}_{\mathrm{DISTS_{MV}}}$ to encourage better style transfer and texture synthesis across volumes. The proposed framework is shown in Fig. \ref{overview}.

\subsection{Conditional Least Squares GAN}
We formulate the highly ill-posed problem of cross-volume synthesis as an image-to-image translation of a patch $\mathbf{x} \in \mathbb{R}^{\mathrm{1 \times H \times W \times D}}$ from the domain of $\boldsymbol{I}_{\mathrm{TLC}}$ patches to a patch $\mathbf{y} \in \mathbb{R}^{\mathrm{1 \times H \times W \times D}}$ within the domain of $\boldsymbol{I}_{\mathrm{RV}}$ patches. The cross entropy-based adversarial feedback in (\ref{cgan}) may cause vanishing gradients that result in highly unstable GAN training. Instead, we used the least squares GAN (LSGAN) framework to stabilize GAN training~\cite{mao2017least}. Given a dataset $\mathcal{Q} = \{(\mathbf{x}_i, \mathbf{y}_i)\}_{i = 1}^{N}$, we trained our framework using the following LSGAN objective:
\begin{equation}
\label{lsgan}
    \mathcal{L}_{\mathrm{LSGAN}} =  -\:\mathbb{E}_{\mathbf{x},\: \mathbf{y}}[(\mathcal{D}(\mathbf{x}, \mathbf{y}) - 1)^{2}] - \mathbb{E}_{\mathbf{x}}[\mathcal{D}(\mathbf{x}, \mathcal{G}(\mathbf{x}))^{2}].    
\end{equation}
To further encourage voxelwise consistency, we proposed to use a coarse-to-fine set of ground-truth labels for deeply supervising the decoder of our generator (see Fig. \ref{overview}). These labels were generated by downsampling the target image patch $\mathbf{y}$ (from the deformed RV image -- $\boldsymbol{I}_{\mathrm{RV}}$) at three different resolutions of the decoder. We used nearest neighbor interpolation for downsampling the image patch $\mathbf{y}$ to generate multiresolution ground-truth labels for supervising our decoder. These labels were used to compute  $\ell_1$ - distance (MR loss or $\mathcal{L}_{\mathrm{MR}}$) at the last three decoder layers, as shown in Fig. \ref{overview}. For last three layers of the decoder, the loss at each resolution $\mathcal{P} = \{\frac{p}{4}, \frac{p}{2}, p\}$ was given as:
\begin{equation}
\label{loss_MR}
\mathcal{L}_{\mathrm{MR}}(\mathbf{y},\: \mathbf{\hat{y}}) = \sum_{i \in \mathcal{P}}\mathbb{E}[\lVert \mathbf{y}_i - \mathbf{\hat{y}}_i \rVert_{1}],
\end{equation}
where $\mathbf{\hat{y}}_{i}$ for $i \in \mathcal{P}$ denoted the generated samples obtained from different layers of the decoder at multiple resolutions (see Fig. \ref{overview}). Similarly, $\mathbf{y}_{i}$ represented the multiresolution ground-truth labels obtained by nearest neighbor interpolation. The MR loss operated on these ground-truth labels and synthetic representations at three different resolutions of the generator decoder, as shown in Fig. \ref{overview}. Before computing the MR loss, we used pointwise 3D convolutions (1 $\times$ 1 $\times$ 1) to align coarse-to-fine tensors across the channel dimensions. The loss for our conditional GAN generator can thus be extended as:
\begin{equation}
    \label{lsgan_G}
    \mathcal{L}_{\mathcal{G}}(\mathbf{x},\: \mathbf{y}) = \underbrace{\mathcal{L}_{\mathrm{MSE}}(\mathcal{D}(\mathbf{x},\: \mathcal{G}(\mathbf{x})), 1)}_\text{least squares adversarial loss~} + \underbrace{\lambda \mathcal{L}_{\mathrm{MR}}(\mathbf{y},\: \mathbf{\mathcal{G}(x)})}_\text{MR $\ell_1$ -- distance~},
\end{equation}
where the first term in (\ref{lsgan_G}) denotes the least squares adversarial loss minimized by the generator. We restricted the loss computation from the final layer to lung segmentation regions for further improving voxel consistency and added it as another term in the loss function.
\subsection{SwinSEER Architecture -- LungViT Generator}
\subsubsection{Hierarchical Shifted-Window Transformer-Based Encoder}
\label{encoder_arch}
We used a multiscale, shifted-window transformer (Swin Transformer) encoder for LungViT (see Fig. \ref{gen}). The encoder backbone consisted of four stages, $\mathrm{\{S_1, S_2, S_3, S_4\}}$, with multiple Swin Transformer blocks~\cite{liu2021swin}. We constructed a hierarchical pyramid of feature maps within the encoder, of size $\mathrm{H \times W \times D} \to \mathrm{\frac{H}{32} \times \frac{D}{32} \times \frac{D}{32}}$, for enhancing image synthesis by learning multiresolution details (see Fig. \ref{gen}). To achieve this, we utilized a patch merging module in each block. Typically, ViTs operate on feature maps at a single resolution, leading to quadratic complexity in computing global self-attention with increasing image size. By using hierarchical representations through patch merging, we were able to perform multi-head self-attention (MSA) within a local patch window at each stage. This enabled Swin Transformer to compute self-attention within a limited window while maintaining linear complexity relative to the image size. A challenge with computing MSA within local windows, however, is that it may not be able to capture relationships between them. This was addressed by following up windowed MSA (W-MSA) with another transformer block with shifted-window MSA (SW-MSA). Two consecutive Swin Transformer blocks with W-MSA and SW-MSA are shown in Fig. \ref{gen}, the computations of which can be expressed as: 
\begin{equation}
    \begin{split}
        \hat{\mathbf{z}}^{l}     &= \text{W-MSA} (\text{LN}(\mathbf{z}^{l - 1}))       + \mathbf{z}^{l - 1}, \\
              \mathbf{z}^{l}     &= \text{MLP}   (\text{LN}(\hat{\mathbf{z}}^{l}))     + \hat{\mathbf{z}}^{l}, \\
        \hat{\mathbf{z}}^{l + 1} &= \text{SW-MSA}(\text{LN}(\mathbf{z}^{l}))           + \mathbf{z}^{l}, \\
              \mathbf{z}^{l + 1} &= \text{MLP}   (\text{LN}(\hat{\mathbf{z}}^{l + 1})) + \hat{\mathbf{z}}^{l + 1}, \\
    \end{split}   
\end{equation}
where $\mathbf{z}^{l}$ and $\mathbf{z}^{l + 1}$ denoted outputs from the regular (W-MSA) and shifted window (SW-MSA) multi-head self-attention, respectively; LN denoted layer normalization and MLP denoted multilayer perceptron. We computed MSA with relative positional bias $\mathrm{B} \in \mathbb{R}^{\mathrm{W^2} \times \mathrm{W^2} \times \mathrm{W^2}}$ for each head using the following expression:
\begin{equation}
\label{msa}
\mathrm{Attention(Q, K, V) = Softmax\left(\frac{QK^T}{\sqrt{d}} + B\right)V},
\end{equation}
where Q, K, V represented query, key, and value tensors, $\mathrm{d}$ indicated size of the queries and keys, and W denoted window size. 

The input to the encoder $\mathbf{x}$ with size $\mathrm{C \times H \times W \times D}$ was mapped to $f$ features by a pointwise 3D convolution operator $\mathcal{M}_1$ and was subsequently partitioned into smaller patches with size $\mathrm{\frac{H}{2} \times \frac{D}{2} \times \frac{D}{2}}$ by a patch partitioning block (as shown in Fig. \ref{gen}). This was processed through four multiresolution stages with Swin Transformer blocks. For all stages, we used 2 Swin Transformer blocks, except in $\mathrm{S}_2$, where we used 4 Swin Transformer blocks. The feature maps were gradually increased to $32f$ for the bottleneck layer, as shown in Fig. \ref{gen}. The input to the encoder was a single-channel patch of size $\mathrm{128 \times 128 \times 128}$.

\subsubsection{Convolutional Decoder with Squeeze-and-Excitation Blocks}
The overall architecture of LungViT generator was a variation of the SwinUNETR architecture~\cite{hatamizadeh2021swin} that used skip connections from multiple blocks of the Swin Transformer encoder to the convolutional decoder (see Fig. \ref{gen}). However, the SwinUNETR decoder fails to capture the relationship between distinct patch features, which are acquired from various encoder blocks and are arranged along the channel dimension. We proposed to enhance the expressiveness of the LungViT decoder by using squeeze-and-excitation blocks (SE blocks)~\cite{hu2018squeeze} for computing channel attention at each resolution of the decoder (see Fig. \ref{gen}). We used four SE blocks in the LungViT decoder and its structure is shown in Fig. \ref{gen}. The SE block works by \textit{squeezing} the spatial dimension of a feature map $\mathcal{F} \in \mathbb{R}^{\mathrm{C \times H \times W \times D}}$ to a vector $\mathbf{z} \in \mathbb{R}^{\mathrm{C}}$ through global average pooling (GAP), such that each element of $\mathbf{z}$, at a location $c$, is computed as:
\begin{equation}
\label{gap}
z_c = \frac{1}{H \times W \times D}\sum_{i = 1}^{H}\sum_{j = 1}^{W}\sum_{k = 1}^{D}\mathcal{F}_c(i, j, k).
\end{equation}
The next step involves \textit{excitation} by passing $\mathbf{z}$ through non-linearities $\delta$ and finally computing the channelwise attention $\mathbf{a}$ by sigmoid activation function $\sigma$:
\begin{equation}
\label{se_func}
\mathbf{a} = \sigma(\mathbf{W}_2 \delta(\mathbf{W}_1 \mathbf{z})) \in \mathbb{R}^{\mathrm{C}},
\end{equation}
where $\mathbf{W}_1 \in \mathbb{R}^{\mathrm{\frac{C}{r} \times C}}$ and $\mathbf{W}_2 \in \mathbb{R}^{\mathrm{C \times \frac{C}{r}}}$ denoted weights for FC layers shown in Fig. \ref{gen}, and $\delta$ denoted ReLU activation function. We used a reduction factor $r = 2$ The final output of the SE block was given by the elementwise multiplication of the channel attention value at location c and the feature map $\mathcal{F}_c \in \mathbb{R}^{\mathrm{H \times W \times D}}$:
\begin{equation}
\label{se_final}
\mathcal{F}^{\prime}_c = a_c\mathcal{F}_c.
\end{equation}

The decoder was made up of five transposed convolution blocks $\{\mathrm{U}_1 \dots \mathrm{U}_5\}$, as shown in Fig. \ref{gen}. We obtained outputs from the last three layers of the decoder for computing multiresolution voxel consistency loss. At every output layer, we used pointwise convolutions $\{\mathcal{M}_2, \mathcal{M}_3, \mathcal{M}_4\}$ for reducing the channel dimensions of the three multiresolution outputs to 1. The final output of the decoder was a single-channel RV volume patch of size $\mathrm{128 \times 128 \times 128}$. During inference, the larger 3D CT volumes, of size 320 $\times$ 320 $\times$ 320, were generated by taking the mean between the overlapping patches extracted with an overlap of 32.

We used a Markovian PatchGAN discriminator in this work~\cite{isola2017image}. Instead of providing adversarial feedback over the entire image, the PatchGAN discriminator attempts to classify image patches, where the size of each patch is smaller than the overall spatial dimension of the input tensor. For input tensor of size 128 $\times $ 128 $\times$ 128, we set the patch size to 8 $\times $ 8 $\times$ 8. Such a patch-based discriminator models the entire image as a Markov random field and helps capture high-resolution details present in smaller patches. We also investigated the impact of using batch normalization and LeakyReLU activation function. These changes, however, did not provide significant improvements in model performance. 

\subsection{Multiview Deep Image Structure and Texture Similarity}
Neural style and texture transfer approaches typically rely on high-dimensional projections of images from large pre-trained models, trained for higher level tasks such as image classification and object detection. The distance between these internal network activations is used to quantify perceptual similarity between two samples. The most commonly used perceptual similarity metric is the learned perceptual image patch similarity, LPIPS~\cite{zhang2018unreasonable}. To compute LPIPS between two slices $\mathbf{s}$, $\hat{\mathbf{s}} \in \mathbb{R}^{\mathrm{C \times H \times W}}$, two sets of high-dimensional embeddings, $\mathcal{H}_{\mathbf{s}}$ and $\mathcal{H}_{\hat{\mathbf{s}}}$, are first extracted from a pre-trained VGG-16 followed by computing $\ell_2$-distance between them. The $\ell_p$-norm based distance functions, however, fail to account for spatial distribution of intensities within an image and therefore may not be suitable for characterizing textural and structural differences between slices. 

\begin{figure}[!t]
\centering
\includegraphics[width=0.45\textwidth]{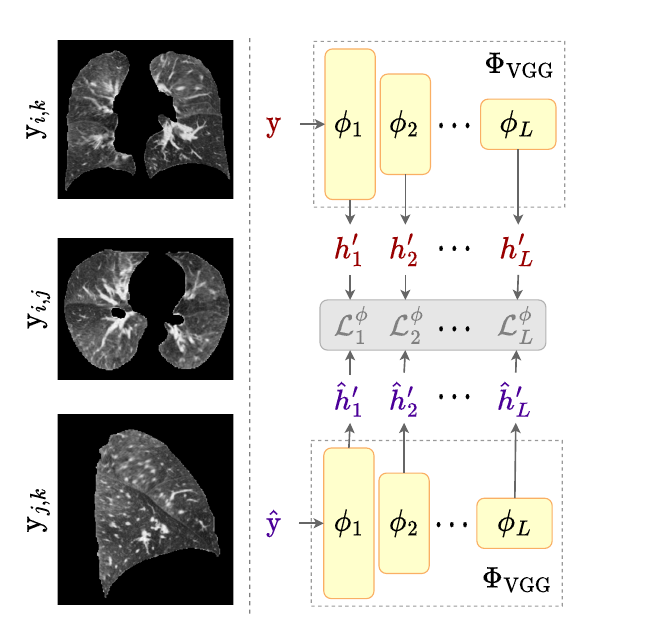}
\vspace{-0.20in}
\caption{Multiview texture similarity module -- DISTS\textsubscript{MV}. A pre-trained VGG - 16 is used to embed one of the three 2D views at a time into a higher-dimensional space. The slice embeddings $\mathcal{H}_{\mathbf{s}}$ and $\mathcal{H}_{\hat{\mathbf{s}}}$ are then used to compute the DISTS measure for each view. A similar approach is used for other two views and the overall multiview texture similarity, DISTS\textsubscript{MV}, is then aggregated.}
\label{distsmv}
\end{figure}

We used the deep image structure and texture similarity (DISTS)~\cite{ding2020image} to characterize both structural and textural differences between the embeddings. For that, we used the formulation proposed by the structural similarity metric (SSIM)~\cite{wang2004image}. Let $\mathcal{H}_{\mathbf{s}} = \{h_i\}_{i = 0}^{L}$ and $\mathcal{H}_{\hat{\mathbf{s}}} = \{\hat{h_i}\}_{i = 0}^{L}$ be two sets of embeddings obtained from a pre-trained VGG-16 model, where $L$ is the number of layers, $i = 0$ denotes original image, and $\{h\}_{i = 1}^{L}$ are the feature maps obtained from different layers of the network. The VGG-16 model was pre-trained on the large-scale hierarchical image repository called the ImageNet~\cite{deng2009imagenet}. We used global means of slice embeddings, $\mu^{i}_{\mathbf{s}}$ and $\mu^{i}_{\hat{\mathbf{s}}}$, to define a quality measure for texture $l(h_i, \hat{h}_i)$ and global correlations of slice embeddings, $\sigma^{i}_{\mathbf{s}}$ and $\sigma^{i}_{\hat{\mathbf{s}}}$, for defining the quality measure for structure $s(h_i, \hat{h}_i)$, defined as:
\begin{equation}
\label{texture}
l(h_i, \hat{h}_i) = \frac{2\mu^{i}_{\mathbf{s}} \mu^{i}_{\hat{\mathbf{s}}} + c_1}{(\mu^{i}_{\mathbf{s}})^2 + (\mu^{i}_{\hat{\mathbf{s}}})^2 + c_1},
\end{equation}

\begin{equation}
\label{structure}
s(h_i, \hat{h}_i) = \frac{2\sigma^{i}_{\mathbf{s}\hat{\mathbf{s}}} + c_2}{(\sigma^{i}_{\mathbf{s}})^2 + (\sigma^{i}_{\hat{\mathbf{s}}})^2 + c_2},
\end{equation}
where $\sigma^{i}_{\mathbf{s}\hat{\mathbf{s}}}$ denotes global covariance between each $h_i$ and $\hat{h}_i$, and $c_1, c_2$ are small constants added to avoid numerical instability. The DISTS loss was then defined by combining (\ref{texture}) and (\ref{structure}) as:

\begin{equation}
\label{dists}
\mathcal{L}_{\mathrm{DISTS}}(\mathbf{s}, \hat{\mathbf{s}}) = 1 - \sum_{i = 0}^{L}(\alpha_il(h_i, \hat{h}_i) + \beta_is(h_i, \hat{h}_i)),
\end{equation}
where $\alpha_i$ and $\beta_i$ were positive learnable weights such that $\sum_{i=0}^{L}(\alpha_i + \beta_i) = 1$. We extended the slice-based objective in (\ref{dists}) to multiple views of a medical image volume. For real $\mathbf{y}$ and synthetic patches $\mathbf{\hat{y}} = \mathcal{G}(\mathbf{x})$ patches, where each $\mathbf{y}$, $\mathbf{\hat{y}} \in \mathbb{R}^{\mathrm{1 \times H \times W \times D}}$, we transposed each volume to compute DISTS across all three dimensions -- axial, coronal and sagittal (shown in Fig. \ref{distsmv}). For computing the similarity across each view, we grouped two dimensions of the 3D patch to independently represent axial ($\mathbf{y}_{i, j}$), coronal ($\mathbf{y}_{i, k}$), sagittal ($\mathbf{y}_{j, k}$) slices, as shown in Fig. \ref{distsmv}. The multiview texture similarity (DISTS\textsubscript{MV}) between two volumes $\mathbf{y}$ and $\mathbf{\hat{y}}$ was thus defined as:
\begin{equation}
\label{mvlpips}
    \begin{split}
    \mathcal{L}_{\mathrm{DISTS_{MV}}}(\mathbf{y}, \mathbf{\hat{y}}) &=  \underbrace{\mathcal{L}(\mathbf{y}_{i, j}, \mathbf{\hat{y}}_{i, j})}_\text{axial~}  + \underbrace{\mathcal{L}(\mathbf{y}_{i, k}, \mathbf{\hat{y}}_{i, k})}_\text{coronal~} \: \\ &+  \underbrace{\mathcal{L}(\mathbf{y}_{j, k}, \mathbf{\hat{y}}_{j, k})}_\text{sagittal~},
    \end{split}
\end{equation}
where $\mathcal{L}(\cdot, \cdot)$ denoted $\mathcal{L}_{\mathrm{DISTS}}(\cdot, \cdot)$ in (\ref{mvlpips}) and each term corresponded to a different view of the volume. This simple extension enabled texture transfer at a volumetric level. The overall cost function minimized by the generator for the proposed framework is:
\begin{equation}
\label{obj_l2l}
\begin{split}
\mathcal{L}_{\mathcal{G}}(\mathbf{x},\: \mathbf{y}) &= \underbrace{\mathcal{L}_{\scaleto{\mathrm{ADV}}{4pt}}(\mathcal{D}(\mathbf{x},\: \mathcal{G}(\mathbf{x})))}_\text{least squares adversarial loss~}   + \underbrace{\lambda_1 \mathcal{L}_{\scaleto{\mathrm{MR}}{4pt}}(\mathbf{y}, \mathcal{G}(\mathbf{x}))}_\text{MR $\ell_1$ -- distance~} \: \\
&+ \underbrace{\lambda_2\mathcal{L}_{\scaleto{\mathrm{DISTS_{MV}}}{4pt}}(\mathbf{y}, \mathcal{G}(\mathbf{x}))}_\text{multiview texture similarity~}.
\end{split}
\end{equation}

\subsection{Context-awareness via ensembling cascade}
\begin{figure}[!t]
\centering
\includegraphics[scale=0.85]{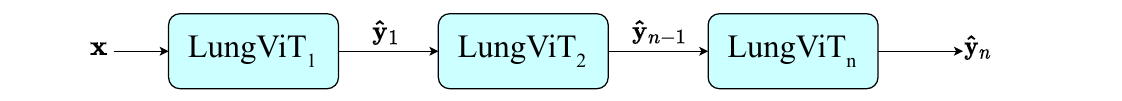}
\centering
\vspace{-0.15in}
\caption{Ensembling by cascading (EC) framework for dense regression where multiple LungViT frameworks are trained, and each subsequent model is conditioned on the output from its precursor model, until a refined output $\mathbf{\hat{y}}_n$ is obtained.}
\label{ac}
\end{figure}

Patch-based generative frameworks often suffer from a lack of global contextual information that is useful for image synthesis. We address this problem by training an ensemble of LungViT frameworks, each conditioned on synthetic outputs from the preceding model, as shown in Fig.~\ref{ac}. Such iterative conditioning helps the successive models to better capture the overall context thereby iteratively improving patch synthesis. It is important to note that each successive model within the ensembling framework did not share weights with the previous model and we used a cascade of two models for this study. Since each model is conditioned on the output from a similar pre-trained predecessor, the modeling framework forms an ensembling cascade called the LungViT\textsubscript{EC}.  

\begin{figure*}[!t]
\centering
\includegraphics[width=\textwidth]{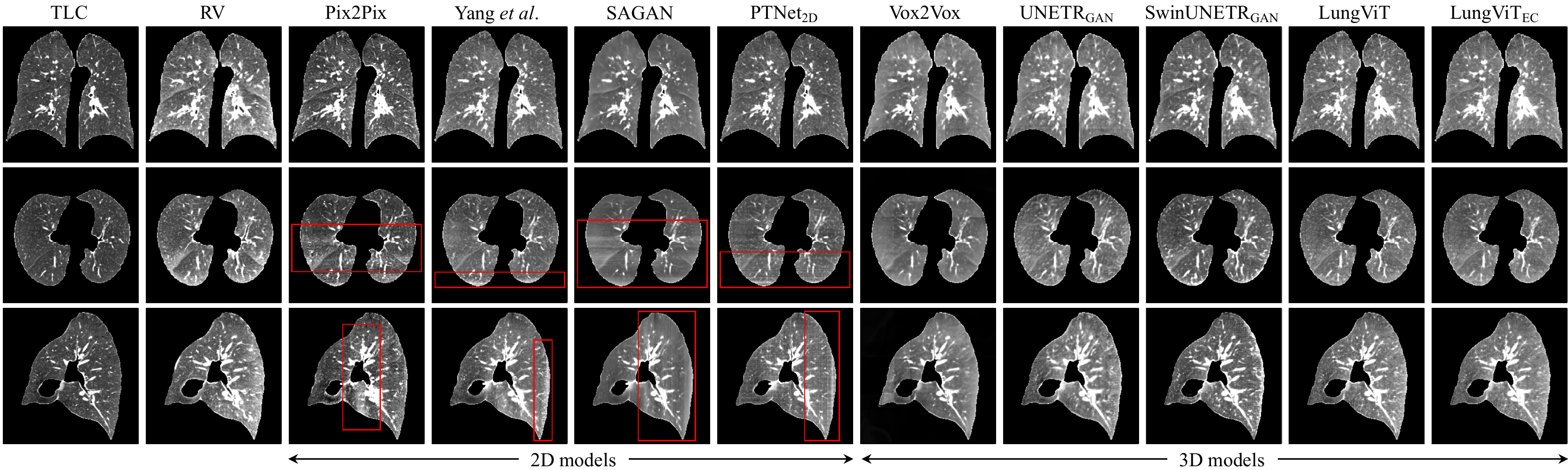}
\vspace{-0.27in}
\caption{Visual comparison of representative samples shown across coronal, axial, and sagittal views from the same subject. Deformed RV image (ground truth in this case) is denoted by RV. We compare results from LungViT with four state-of-the-art slice-based methods, Pix2Pix~\cite{isola2017image}, Yang \textit{et al}.~\cite{yang2018low}, SAGAN~\cite{zhang2019self}, and PTNet~\cite{zhang2022ptnet3d}, and three different volumetric methods including, Vox2Vox~\cite{johnson2019conditional}, UNETR~\cite{hatamizadeh2022unetr}, and SwinUNETR~\cite{hatamizadeh2021swin}. Since UNETR and SwinUNETR were trained adversarially, we denote them as UNETR\textsubscript{GAN} and SwinUNETR\textsubscript{GAN}, respectively. All 2D methods showed discontinuities across axial and sagittal views (shown in red), since they were trained on mid-coronal slices.}
\label{qual}
\end{figure*}

\section{Experimental Details}
\subsection{Dataset and Preprocessing}
For training and evaluating our models, we utilized CT data from the SubPopulations and Intermediate Outcome Measures in COPD Study (SPIROMICS)~\cite{couper2014design}. SPIROMICS is a multi-center, prospective cohort study that acquired breathhold CT scans for every participant at two different lung volumes -- total lung capacity (TLC) and residual volume (RV). The original resolution of the chest CT scans was approximately 0.6 $\times$ 0.6 $\times$ 0.5 mm$^3$, and the image size was 512 $\times$ 512 per slice, with 500 to 600 slices per volume~\cite{sieren2016spiromics}. SPIROMICS acquired scans from subjects with varying degrees of disease severity, defined by the Global Initiative for Chronic Obstructive Lung Disease (GOLD)~\cite{vestbo2013global}. GOLD 1 to GOLD 4 denotes mild to severe COPD, while asymptomatic smokers were grouped in GOLD 0. A small subset of normal individuals who never smoked was also included for analysis (see Table \ref{t_1}).

Before training, we registered RV volumes to the corresponding TLC volumes using a mass-preserving DIR method~\cite{yin2009mass, gorbunova2012mass,cao2010regularized}. The transformation between two volumes was parameterized using cubic BSpline interpolation and was determined using a multiresolution strategy from Yin \textit{et al}~\cite{yin2009mass}. The overall cost function for the DIR method included a tissue mass-preserving component for matching image intensities, a structural constraint to encourage better alignment between fissure structures, and a regularization term on displacement vector field to ensure a smooth transformation~\cite{yin2009mass, gorbunova2012mass,cao2010regularized}. Pre-registered image volumes at TLC and RV were then resampled isotropically to a resolution of 1 $\times$ 1 $\times$ 1 mm$^3$. To remove outliers arising due to calcification or metal artifacts, intensity values were clipped to the interval [-1024, 1024] Hounsfield units (HU). The image volumes were then cropped to the bounding box containing the union of the lung regions of the inspiration and expiration. A multi-resolution convolutional neural network was used to segment the lung regions~\cite{gerard2020multi}. We further normalized the image intensities between -1 and 1. Disjoint training and testing sets had 1055 and 512 subjects, spanning the GOLD spectrum of disease severity, as illustrated in Table \ref{t_1}.

\begin{table}[htbp]
    \centering
    \caption{Distribution of COPD severity, defined by GOLD~\cite{vestbo2013global}, across disjoint training and testing datasets.}
    \vspace{-0.1in}
    \begin{tabularx}{0.48\textwidth}{l c c}
    \toprule
      & \textbf{Training} & \textbf{Testing} \\
    \midrule
    Individuals who never smoked & 154 & 59 \\
    GOLD 0       & 200  & 100 \\
    GOLD 1       & 200  & 99  \\
    GOLD 2       & 200  & 100 \\
    GOLD 3       & 200  & 100 \\
    GOLD 4       & 133  & 54  \\
    \midrule
    Total        & 1055 & 512 \\
    \bottomrule
\end{tabularx}
    \label{t_1}
\end{table}

\begin{table*}[h]
    \small
    \centering
    \caption{Quantitative performance evaluation of LungViT across 512 subjects from the SPIROMICS cohort, in contrast to various state-of-the-art 2D and 3D generative frameworks. Performance was evaluated using PSNR, SSIM, LPIPS, FID, NMSE, \\ and mean absolute error between ground truth and predicted fsad values. Wilcoxon's signed rank test was used to  \\assess differences between the means of all evaluation metrics with LungViT\textsubscript{EC} selected as the reference and \\ $^{\ast}: p < 0.05$, $^{\ast\ast}: p < 0.01$, $^{\ast\ast\ast}: p < 0.001$, and $^{\dagger}: p >= 0.05$. The best metric values are highlighted in \textbf{black} \\ and the second best values are highlighted in \textcolor{BlueViolet}{\textbf{blue}}.}
    \vspace{-0.1in}
    \begin{tabularx}{0.98\textwidth}{l l l l l l l}
    \toprule
      & \textbf{PSNR (dB)}\textsubscript{$\uparrow$} & \textbf{SSIM}\textsubscript{$\uparrow$} & \textbf{LPIPS}\textsubscript{2.5D}\textsubscript{$\downarrow$} & \textbf{FID}\textsubscript{2.5D}\textsubscript{$\downarrow$}  & \textbf{NMSE ($\%$)}\textsubscript{$\downarrow$} & \textbf{fSAD}\textsubscript{MAE}\textsubscript{$\downarrow$} \\
    \midrule
    Pix2Pix & 22.42 $\pm$ 1.39$^{\ast\ast\ast}$ & 0.795 $\pm$ 0.028$^{\ast\ast\ast}$ & 0.123 $\pm$ 0.019$^{\ast\ast\ast}$ & 2.06 $\pm$ 0.47$^{\ast\ast\ast}$ & 4.94 $\pm$ 3.41$^{\ast\ast\ast}$ & 10.15 $\pm$ 6.89$^{\ast\ast\ast}$  \\
    Yang \textit{et al}. & 24.17 $\pm$ 2.32$^{\ast\ast\ast}$ & 0.833 $\pm$ 0.039$^{\ast\ast\ast}$ & 0.101 $\pm$ 0.020$^{\ast\ast\ast}$ & 1.62 $\pm$ 0.44$^{\ast\ast\ast}$ & 3.66 $\pm$ 2.87$^{\ast\ast\ast}$ & 6.09 $\pm$ 6.43$^{\dagger}$ \\
    SAGAN & \textcolor{BlueViolet}{\textbf{24.58}} $\pm$ \textcolor{BlueViolet}{\textbf{2.67}}$^{\ast\ast\ast}$ & \textbf{0.845} $\pm$ \textbf{0.039}$^{\ast\ast\ast}$ & 0.145 $\pm$ 0.023$^{\ast\ast\ast}$ & 3.22 $\pm$ 0.84$^{\ast\ast\ast}$ & 3.52 $\pm$ 3.04$^{\ast\ast\ast}$ & 6.31 $\pm$ 6.68$^{\dagger}$  \\
    PTNet & 24.16 $\pm$ 1.39$^{\ast\ast\ast}$ & 0.832 $\pm$ 0.040$^{\ast\ast\ast}$ & 0.114 $\pm$ 0.022$^{\ast\ast\ast}$ & 2.49 $\pm$ 0.78$^{\ast\ast\ast}$ & 3.66 $\pm$ 2.83$^{\ast\ast\ast}$ & 6.63 $\pm$ 7.22$^{\ast}$  \\
    \midrule
    Vox2Vox & 24.04 $\pm$ 1.95$^{\ast\ast\ast}$ & 0.819 $\pm$ 0.032$^{\ast\ast\ast}$ & 0.159 $\pm$ 0.019$^{\ast\ast\ast}$ & 3.48 $\pm$ 0.48$^{\ast\ast\ast}$ & 3.63 $\pm$ 2.85$^{\ast\ast\ast}$ & 9.33 $\pm$ 9.22$^{\ast\ast\ast}$ \\
    UNETR\textsubscript{GAN} & 23.93 $\pm$ 2.18$^{\ast\ast\ast}$ & 0.830 $\pm$ 0.035$^{\ast\ast\ast}$ & 0.121 $\pm$ 0.019$^{\ast\ast\ast}$ & 2.34 $\pm$ 0.43$^{\ast\ast\ast}$ & 3.79 $\pm$ 2.89$^{\ast\ast\ast}$ & 6.40 $\pm$ 7.00$^{\dagger}$  \\
    SwinUNETR\textsubscript{GAN} & 23.53 $\pm$ 2.11$^{\ast\ast\ast}$ & 0.823 $\pm$ 0.038$^{\ast\ast\ast}$ & 0.111 $\pm$ 0.019$^{\ast\ast\ast}$ & 1.81 $\pm$ 0.47$^{\ast\ast\ast}$ & 4.09 $\pm$ 2.91$^{\ast\ast\ast}$ & 7.55 $\pm$ 7.27$^{\ast\ast\ast}$  \\
     \midrule
    LungViT & 24.43 $\pm$ 2.42$^{\ast\ast\ast}$ & 0.837 $\pm$ 0.042$^{\ast\ast\ast}$ & \textcolor{BlueViolet}{\textbf{0.096}} $\pm$ \textcolor{BlueViolet}{\textbf{0.019}}$^{\ast\ast\ast}$ & \textbf{1.42} $\pm$ \textbf{0.44}$^{\ast\ast\ast}$ & \textcolor{BlueViolet}{\textbf{3.50}} $\pm$ \textcolor{BlueViolet}{\textbf{2.90}}$^{\ast\ast\ast}$ & \textbf{6.04} $\pm$ \textbf{7.06}$^{\dagger}$  \\
    LungViT\textsubscript{EC} & \textbf{24.75} $\pm$ \textbf{2.47} & \textcolor{BlueViolet}{\textbf{0.842}} $\pm$ \textcolor{BlueViolet}{\textbf{0.040}} & \textbf{0.096} $\pm$ \textbf{0.020} & \textcolor{BlueViolet}{\textbf{1.51}} $\pm$ \textcolor{BlueViolet}{\textbf{0.43}} & \textbf{3.26} $\pm$ \textbf{2.70}  & \textcolor{BlueViolet}{\textbf{6.06}} $\pm$ \textcolor{BlueViolet}{\textbf{7.07}}  \\
    \bottomrule
\end{tabularx}
    \label{r_1_r2}
\end{table*}

\subsection{Model Evaluation}
\subsubsection{Quantitative Evaluation}
We evaluated our models using different voxel-based, structural, and perceptual quantitative metrics. To assess voxelwise differences between the real images $\mathbf{y}$ and synthetic images $\mathbf{\hat{y}} = \mathcal{G}(\mathbf{x})$, we used peak-signal-to-noise-ratio (PSNR) in decibels (dB) and percent normalized mean squared error (NMSE). The voxelwise metrics were computed within the lung region defined by the segmentation mask. PSNR between $\mathbf{y}$ and $\mathbf{\hat{y}}$ was defined as $\mathrm{PSNR}(\mathbf{y}, \mathbf{\hat{y}}) = 20\: \mathrm{log}_{10}\: \frac{\mathrm{max\{\mathbf{y}\}}}{\lVert \mathbf{y} - \mathbf{\hat{y}} \rVert_{\ell_2}}$, where $\mathrm{max\{\mathbf{y}\}}$ was the maximum possible value of image intensities, and NMSE was expressed as $\mathrm{NMSE}(\mathbf{y}, \mathbf{\hat{y}}) = \frac{\lVert \mathbf{y} - \mathbf{\hat{y}} \rVert_{\ell_2}^{2}}{\lVert \mathbf{y} \rVert_{\ell_2}^{2}}$. To assess local structural similarity between the real and synthetic volumes we used the structural similarity index (SSIM)~\cite{wang2004image} stated as $\mathrm{SSIM}(\mathbf{y}, \hat{\mathbf{y}}) = \frac{(2\mu_{\mathbf{y}}\mu_{\mathbf{\hat{y}}} + c_1)(2\sigma_{\mathbf{y}\mathbf{\hat{y}}} + c_2)}{(\mu^{2}_{\mathbf{y}} + \mu^{2}_{\mathbf{\hat{y}}} + c_1)(\sigma^{2}_{\mathbf{y}} + \sigma^{2}_{\mathbf{\hat{y}}} + c_2)}$, where $\mathbf{y}$ and $\mathbf{\hat{y}}$ indicated real and synthetic 3D volumes, respectively. Here, $\mu$'s denoted means and $\sigma$'s denote standard deviations and a window of size 11 $\times$ 11 $\times$ 11 was used. We also evaluated the perceptual similarity between real and synthetic images using LPIPS~\cite{zhang2018unreasonable} and Fr\'echet Inception Distance (FID)~\cite{heusel2017gans}. To further assess agreement between the means of real and synthetically generated images, we conducted a Bland-Altman analysis~\cite{giavarina2015understanding}.

\subsubsection{Clinical Validation}
Most of the studies involved with the development of generative models tend to ignore clinical validation of the synthetic samples generated by their models. We sought to evaluate the clinical reliability of our samples by computing air trapping and parametric response mapping (PRM)-based emphysema and functional small airways disease (fSAD). These biomarkers, which require matched inspiratory and expiratory image pairs, have been shown to be associated with various disease outcomes in COPD~\cite{galban2012computed}. fSAD was defined as the fraction of voxels between -950 HU and -810 HU on a TLC scan and between -1000 HU and -857 HU on an RV scan, while air trapping was defined as the percent of voxels below -856 HU on an RV scan alone.

\subsubsection{Qualitative Evaluation}
We evaluated visual sample quality by segmenting pulmonary fissures on the real and synthetic images. In doing so, we sought to evaluate whether the synthetic volumes captured the shape and grayscale characteristics of the fissures reliably. Fissure segmentation was performed using FissureNet, a state-of-the-art lung fissure segmentation method~\cite{gerard2018fissurenet}. We also evaluated the spatial distribution of air trapping on real and synthetic scans.

\subsubsection{Out-of-Distribution Testing}
To assess model generalizability, we evaluated our model on CT data obtained from a totally different cohort from the Genetic Epidemiology of COPD study (COPDGene)~\cite{regan2011genetic}. We used CT data from 200 subjects of the COPDGene study that were acquired at the University of Iowa Hospitals and Clinics using a different CT acquisition protocol. Model evaluation on an out-of-distribution cohort was conducted to evaluate model biases towards the validation set. The CT images were processed and registered following a similar method used for the images from SPIROMICS.

\subsubsection{Registration Evaluation}
To assess the performance of image registration method used in this study, we measured the overlap between TLC image masks and deformed RV image masks. The overlap was quantified using three different metrics: Dice similarity coefficient (DSC)~\cite{dice1945measures}, the Jaccard index~\cite{jaccard1912distribution}, and average symmetric surface distance (ASSD) measured in millimeters (mm)~\cite{heimann2009comparison}.

\subsection{Comparison with State-of-the-Art}
We compared the performance of LungViT model with several state-of-the-art 2D and 3D convolutional, self-attention- and transformer-based generative models. 
\subsubsection{Convolutional Generative Models}
We compared model performance with Pix2Pix~\cite{isola2017image}, a slice-based 2D GAN framework for conditional image-to-image translation. We also compared model performance with a volumetric variant of Pix2Pix, the Vox2Vox framework~\cite{johnson2019conditional}. Another convolutional model was the work by Yang \textit{et al}.~\cite{yang2018low}. We compared our model to this work because it was one of the first GAN frameworks to jointly optimize perceptual and norm-based distances for low-dose CT denoising, and several recent generative modeling techniques have been influenced by the ideas presented in this work.

\subsubsection{Self-Attention and Transformer-Based Models}
We also compared model performance with self-attention GAN (SAGAN)~\cite{zhang2019self}. SAGAN uses convolutional self-attention blocks within the decoder to model long-range dependencies between pixels of an image. It also uses spectral normalization as a stability measure for adversarial learning~\cite{zhang2019self}. Our model performance was compared to two hybrid convolutional and transformer-based models, UNETR~\cite{hatamizadeh2022unetr} and SwinUNETR~\cite{hatamizadeh2021swin}, and a recently proposed purely transformer-based generative model called PTNet~\cite{zhang2022ptnet3d}. To improve perceptual performance of transformer models, both UNETR and SwinUNETR were trained adversarially on 3D image patches and were called UNETR\textsubscript{GAN} and SwinUNETR\textsubscript{GAN}. 

\subsubsection{Ablation Study}
We conducted an ablation study to assess the relative contribution of various model components by gradually increasing the model complexity.

\begin{table*}[htbp]
    \small
    \centering
    \caption{Performance evaluation of the LungViT\textsubscript{EC} framework across varying degrees of disease severity, as defined by the Global Initiative for Chronic Obstructive Lung Disease (GOLD)~\cite{vestbo2013global}.}
    \vspace{-0.1in}
    \begin{tabularx}{0.835\textwidth}{l l l l l l}
    \toprule
      & \textbf{PSNR (dB)}\textsubscript{$\uparrow$} & \textbf{SSIM}\textsubscript{$\uparrow$} & \textbf{LPIPS}\textsubscript{2.5D}\textsubscript{$\downarrow$} & \textbf{FID}\textsubscript{2.5D}\textsubscript{$\downarrow$}  & \textbf{NMSE ($\%$)}\textsubscript{$\downarrow$} \\
    \midrule
    Individuals who never smoked & 22.97 $\pm$ 1.45 & 0.820 $\pm$ 0.031 & 0.102 $\pm$ 0.018 & 1.89 $\pm$ 0.35 & 5.24 $\pm$ 2.32  \\
    GOLD 0 & 23.82 $\pm$ 1.78 & 0.832 $\pm$ 0.032 & 0.096 $\pm$ 0.017 & 1.71 $\pm$ 0.38 & 4.28 $\pm$ 3.08 \\
    GOLD 1 & 23.57 $\pm$ 2.36 & 0.821 $\pm$ 0.050 & 0.109 $\pm$ 0.032 & 1.76 $\pm$ 1.00 & 4.20 $\pm$ 3.35 \\
    GOLD 2 & 24.59 $\pm$ 1.89 & 0.842 $\pm$ 0.036 & 0.097 $\pm$ 0.020 & 1.51 $\pm$ 0.28 & 2.93 $\pm$ 1.83 \\
    GOLD 3  & 26.20 $\pm$ 2.15 & 0.859 $\pm$ 0.038 & 0.090 $\pm$ 0.019 & 1.20 $\pm$ 0.30 & 1.74 $\pm$ 1.05 \\
    GOLD 4  & 28.03 $\pm$ 1.91 & 0.887 $\pm$ 0.023 & 0.081 $\pm$ 0.012 & 1.03 $\pm$ 0.27 & 1.08 $\pm$ 0.83 \\
    \bottomrule
\end{tabularx}
    \label{r_2}
\end{table*}

\subsection{Implementation Details}
The LungViT framework consisted of different modules including the generator, discriminator, and a pre-trained VGG-16 model. The framework was implemented using the open source framework PyTorch~\cite{paszke2019pytorch} and MONAI~\cite{cardoso2022monai}, and was trained using a single NVIDIA A100 GPU with a batch size of 4. During training, the entire framework used 75 GBs of GPU memory for end-to-end training. Since the weights of pre-trained VGG-16 model were not optimized by computing the backpropagation gradients, it utilized a smaller fraction of the overall GPU memory. For optimizing the generator and discriminator parameters, we used the Adam optimizer with imbalanced learning rates 0.0002 and 0.00005, respectively. The hyperparameters $\lambda_1$ and $\lambda_2$ were set to 100 for all experiments.

\begin{table}[htbp]
    \centering
    \caption{Quantitative evaluation of the deformable image registration indicated by the overlap between TLC image masks and deformed RV image masks. The overlap was quantified  \\ by the Dice similarity coefficient (DSC), the Jaccard \\ index, and the average symmetric surface distance \\ (ASSD). Registration performance across training \\ and testing sets has also been reported.}
    \vspace{-0.1in}
    \begin{tabularx}{0.42\textwidth}{l c c c}
    \toprule
      & \textbf{DSC} & \textbf{Jaccard} & \textbf{ASSD (mm)}\\
    \midrule
    Training       & 0.988 $\pm$ 0.008  & 0.977 $\pm$ 0.013 & 0.459 $\pm$ 0.375\\
    Testing        & 0.988 $\pm$ 0.009  & 0.976 $\pm$ 0.015 & 0.484 $\pm$ 0.423\\
    \midrule
    Overall        & 0.988 $\pm$ 0.008  & 0.976 $\pm$ 0.014 & 0.467 $\pm$ 0.391 \\
    \bottomrule
\end{tabularx}
    \label{t_4}
\end{table}

\section{Results}
\begin{figure}[!t]
\centering
\includegraphics[width=0.48\textwidth]{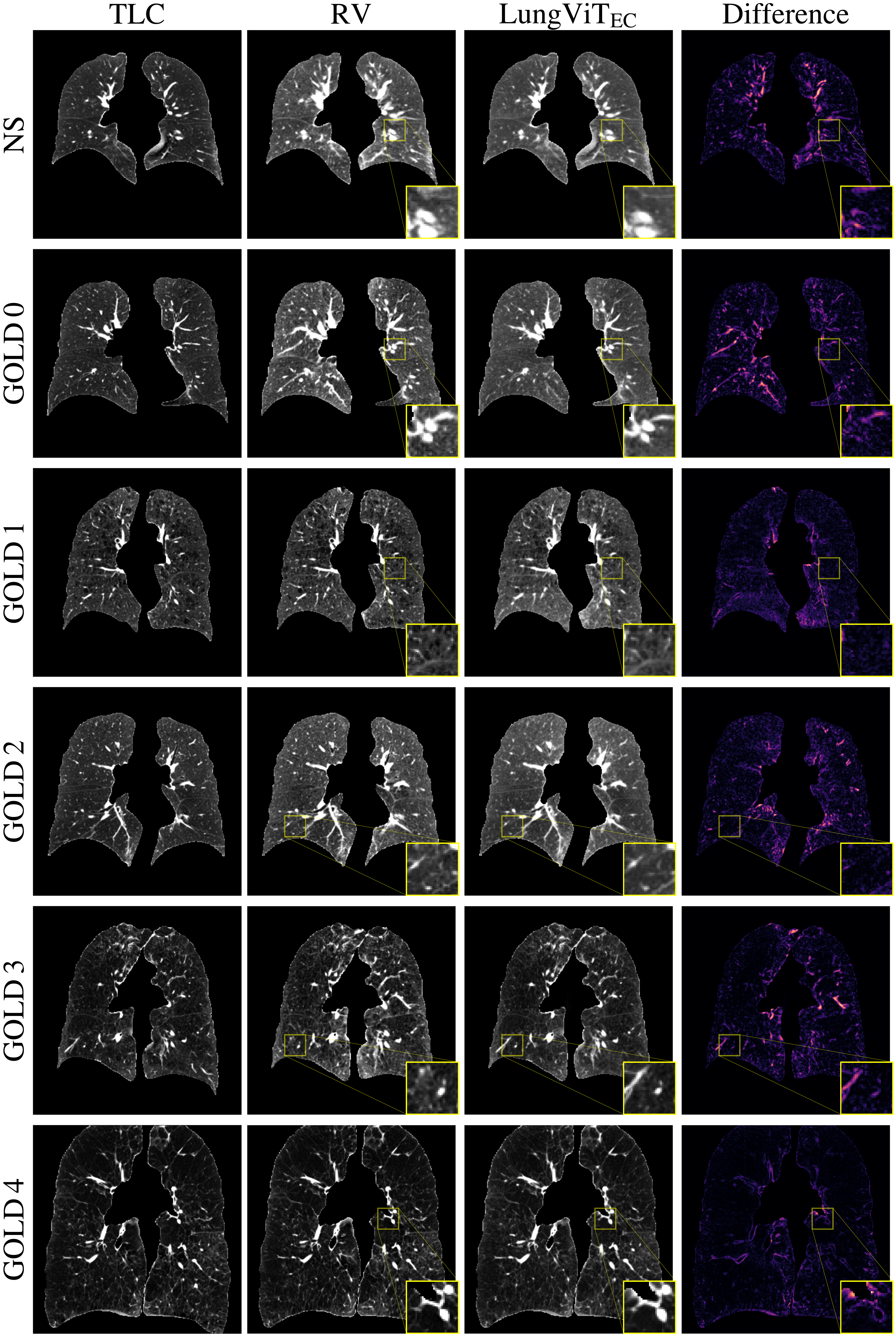}
\vspace{-0.12in}
\caption{Qualitative performance analysis of LungViT\textsubscript{EC} across varying COPD severity. `NS' indicates individuals who never smoked. RV denotes ground truth scans deformed to TLC space.}
\label{q_v_gold}
\end{figure}

We evaluated our model using several qualitative and quantitative measures and compared its performance with various state-of-the-art volumetric and slice-based generative models. Representative synthesis results in multiple cross-sectional views from all models have been shown in Fig. \ref{qual}. All 3D models, including Vox2Vox, UNETR\textsubscript{GAN}, SwinUNETR\textsubscript{GAN}, LungViT, and LungViT\textsubscript{EC}, generated spatially consistent results in all three dimensions. In contrast, the 2D GANs, which were trained on coronal slices, produced discontinuities across axial and sagittal views (highlighted in red boxes in Fig. \ref{qual}). The discontinuities were most prominent for SAGAN and PTNet, while Pix2Pix and the method from Yang \textit{et al}~\cite{yang2018low} showed bands of discontinuities in some regions. On sagittal cross-sections, the discontinuities were observed mostly on the dorsal side of the lung. The Vox2Vox generated blurred images with missing tissue textures, lobar fissures, and bronchovascular bundles (see Fig. \ref{qual}). We also show mid-coronal slices from six different subjects with varying GOLD stages for LungViT\textsubscript{EC} in Fig. \ref{q_v_gold}. The LungViT\textsubscript{EC} was able to capture aeration change patterns, intricate vessel structures, and fissures across all GOLD stages (see Fig. \ref{q_v_gold}).

The LungViT framework, which incorporated DISTS\textsubscript{MV} and SE blocks, showed superior quantitative performance when compared to seven other methods (see Table \ref{r_1_r2}). We observed a significant increase in performance with ensemble cascading (EC), as shown in Table \ref{r_1_r2} for LungViT\textsubscript{EC}. LungViT\textsubscript{EC} achieved an overall PSNR of 24.75 followed by SAGAN with PSNR 24.58 and LungViT with PSNR 24.43. Both LungViT\textsubscript{EC} and LungViT showed high perceptual quality with significantly lower LPIPS (0.0962 and 0.0964) and FID (1.51 and 1.42) values. LungViT\textsubscript{EC} and LungViT also performed superiorly in terms of NMSE (3.26\% and 3.50\%) and fSAD\textsubscript{MAE} (6.06 and 6.04). The highest SSIM was achieved by SAGAN (0.845) followed by LungViT\textsubscript{EC} (0.842) and LungViT (0.837). Wilcoxon's signed ranked test indicated a statistically significant difference in performance between LungViT\textsubscript{EC} and all other models. Quantitative performance of LungViT\textsubscript{EC} across varying levels of disease severity, ranging from GOLD 0 to GOLD 4, is shown in Table \ref{r_2}. Quantitative performance evaluation of the image registration has been reported in Table \ref{t_4}. The overlap between TLC and deformed RV image masks showed good agreement with an overall DSC of 0.988, Jaccard index of 0.967, and ASSD of 0.467mm.

\begin{figure}[!t]
\centering
\includegraphics[width=0.50\textwidth]{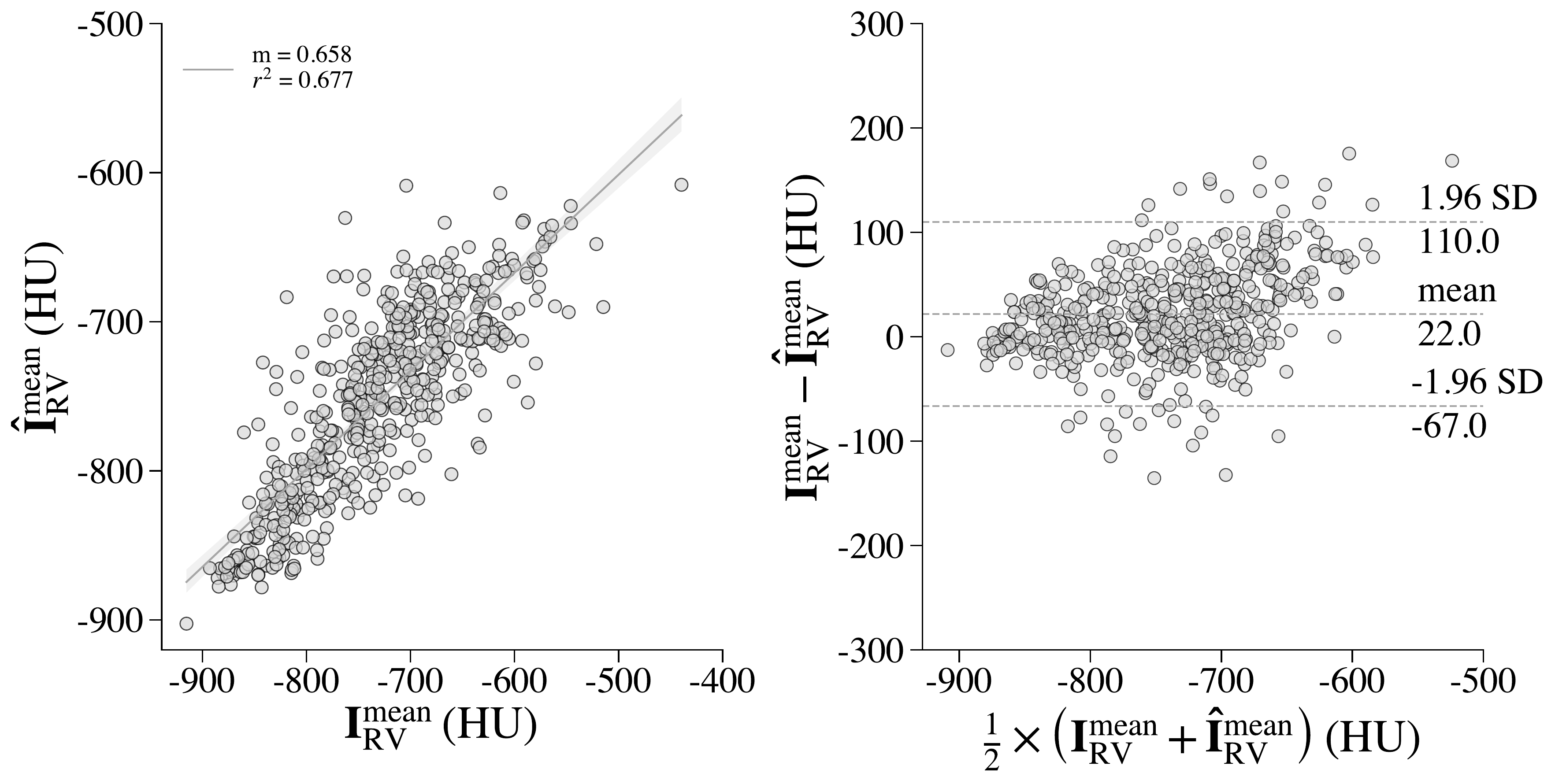}
\vspace{-0.10in}
\caption{Regression analysis with corresponding Bland-Altman plot for comparison between the means of real and synthetic RV scans generated by LungViT\textsubscript{EC}. $\boldsymbol{I}_{\mathrm{RV}}$ denotes ground truth RV scans deformed into TLC image space.}
\label{altman}
\end{figure}

We conducted a regression analysis for LungViT\textsubscript{EC} between the means of real $\boldsymbol{I}^{\mathrm{mean}}_{\mathrm{RV}}$ and synthetically generated RV scans $\hat{\boldsymbol{I}}^{\mathrm{mean}}_{\mathrm{RV}}$ (see Fig. \ref{altman}). The means showed good agreement with an overall coefficient of determination $r^2 = 0.66$. A Bland Altman analysis showed a small bias of 22 HU between the means of real and synthetic RV scans (Fig. \ref{altman}).

\begin{figure}[!t]
\centering
\includegraphics[width=0.50\textwidth]{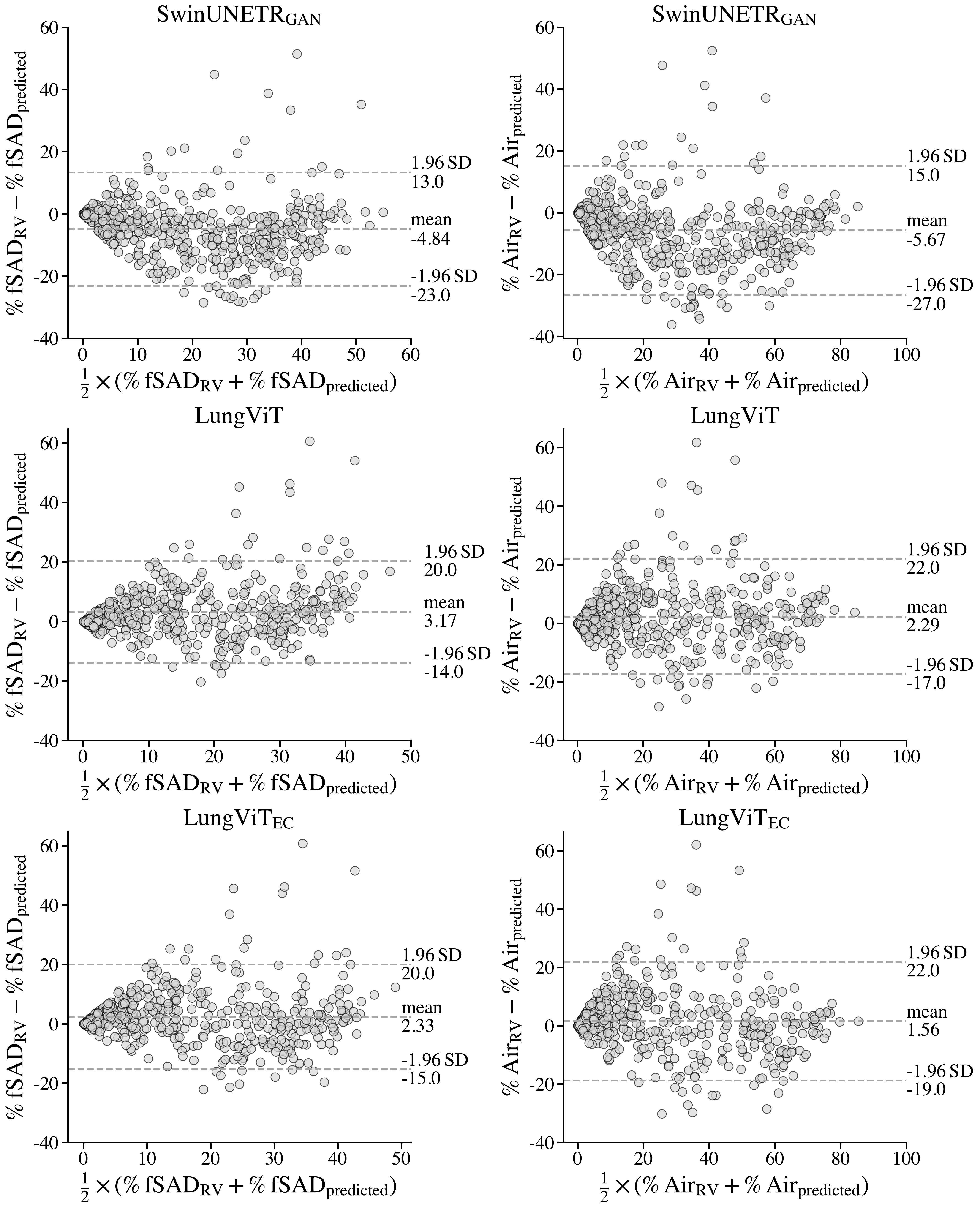}
\vspace{-0.18in}
\caption{Bland Altman analysis for comparing the CT biomarkers ($\%$ fSAD and $\%$ air trapping) derived from real RV volumes (deformed to TLC space) and synthetic RV volumes, generated by SwinUNETR\textsubscript{GAN}, LungViT, and LungViT\textsubscript{EC}. $\%$ fSAD\textsubscript{RV} and $\%$ Air\textsubscript{RV} indicate the ground truth values extracted from the deformed RV image volumes. $\%$ fSAD\textsubscript{predicted} and $\%$ Air\textsubscript{predicted}were the biomarker values extracted from model predicted, synthetic RV images. In each Bland Altman plot, mean difference and associated confidence intervals (defined as $\pm$1.96 SD) are also provided. SD = standard deviation of the difference.}
\label{biomarker}
\end{figure}

We conducted model validation using two well-known imaging biomarkers that are derived from TLC and RV scans for characterizing COPD -- CT density-based air-trapping~\cite{newman1994quantitative} and image registration-based PRM characterization of functional small airways disease (fSAD)~\cite{galban2012computed}. Percent air trapping was defined as the fraction of voxels below -856 HU within the lung region on an RV scan. For computing fSAD, joint histograms of co-registered TLC-RV image pairs were used. fSAD was defined as the fraction of voxels between -950 HU and -810 HU on a TLC scan and between -1000 HU and -857 HU on an RV scan. We assessed differences in  CT biomarkers derived from real and synthetically generated RV volumes by conducting a Bland Altman analysis shown in Fig. \ref{biomarker}. Synthetic RV images generated by LungViT\textsubscript{EC} yielded similar air trapping and fSAD percentages with a minimal overall bias of 2.33\% for fSAD and 1.56\% for air trapping. Similarly, LungViT showed a bias of 3.17\% for fSAD and 2.29\% for air trapping estimation. Both LungViT\textsubscript{EC} and LungViT underestimated fSAD and air trapping. SwinUNETR\textsubscript{GAN} significantly overestimated fSAD and air trapping with negative biases 4.84\% and 5.67\%, respectively. Unlike LungViT\textsubscript{EC} and LungViT, the errors in SwinUNETR\textsubscript{GAN} were not consistent (see Fig. \ref{biomarker}).

\begin{figure}[!t]
\centering
\includegraphics[width=0.50\textwidth]{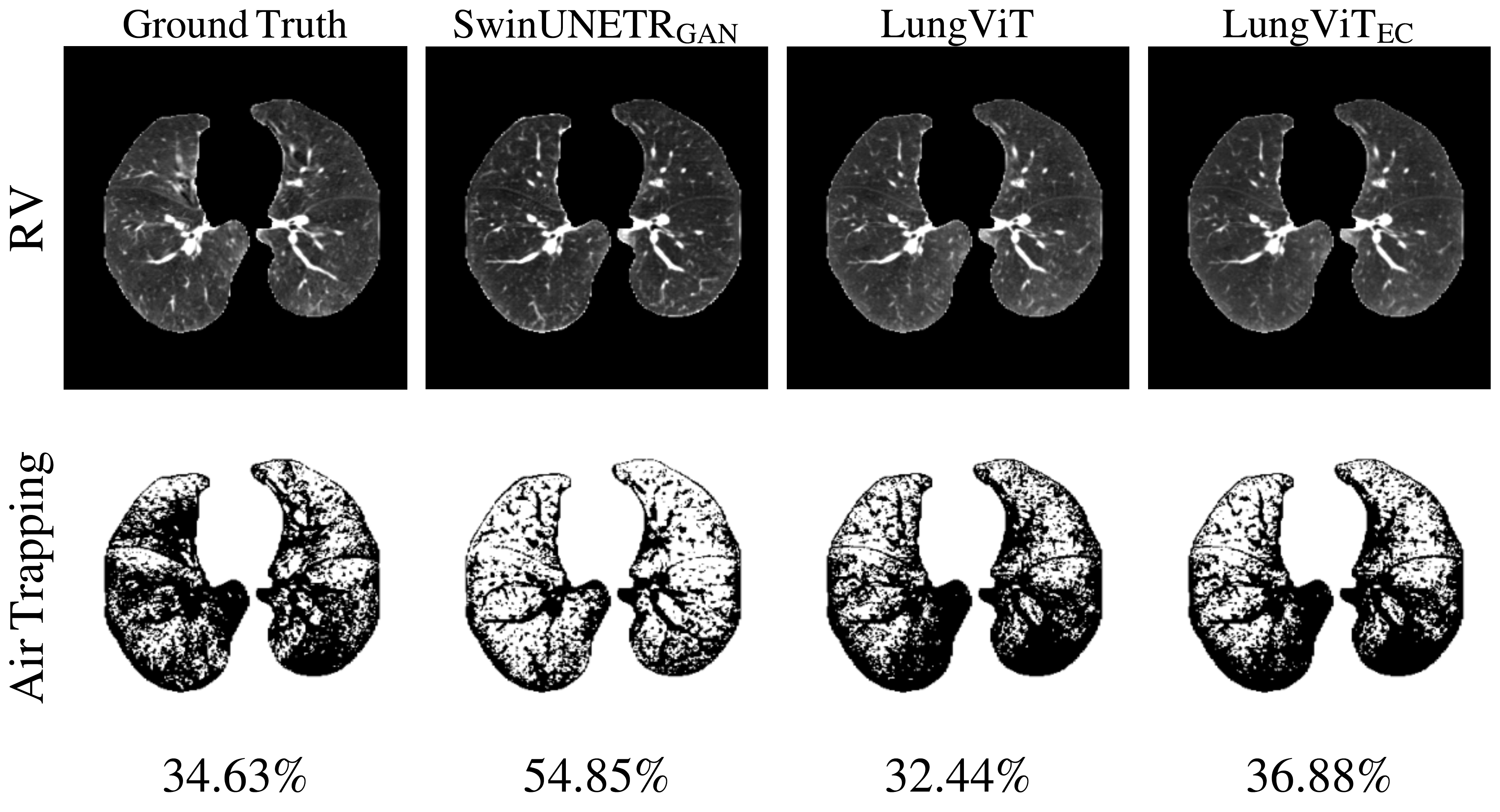}
\vspace{-0.15in}
\caption{Impact of DISTS\textsubscript{MV} on visual the distribution of air trapping. SwinUNETR\textsubscript{GAN}, for instance, overestimated the overall percent air trapping, while the LungViT models performed well at texture synthesis which improved air trapping estimation.}
\label{impact}
\end{figure}

\begin{figure}[!t]
\centering
\includegraphics[width=0.49\textwidth]{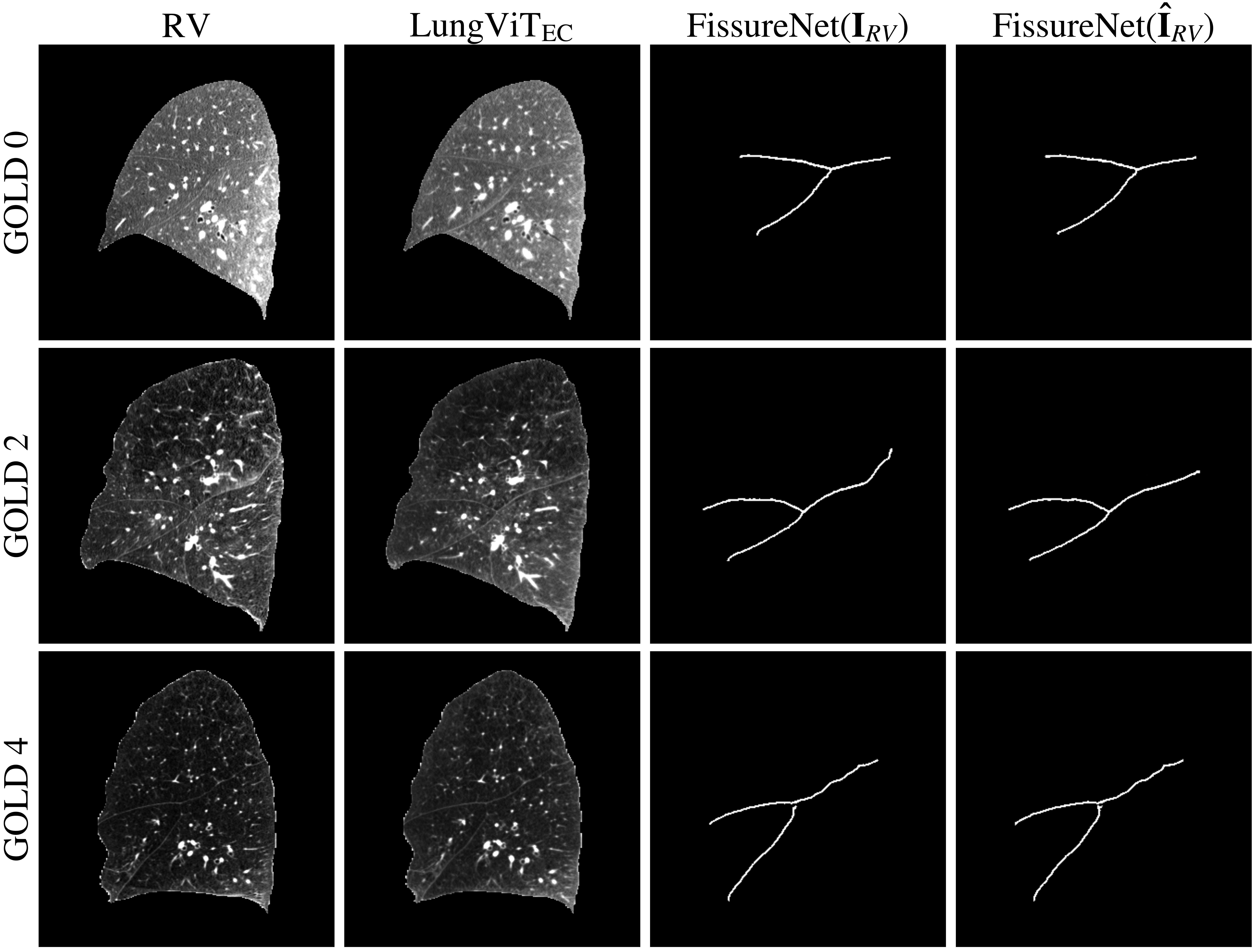}
\vspace{-0.20in}
\caption{Fissure segmentations using a state-of-the-art fissure segmentation model, FissureNet~\cite{gerard2018fissurenet} on real $\boldsymbol{I}_{\mathrm{RV}}$ (deformed or moving RV image) and synthetic $\boldsymbol{\hat{I}}_{\mathrm{RV}}$ samples from LungViT\textsubscript{EC}, shown on sagittal slices.}
\label{fissures}
\end{figure}

In Fig. \ref{impact}, we show the effectiveness of the DISTS\textsubscript{MV} module towards capturing the spatial distribution of air trapping (shown on axial slices in comparison to SwinUNETR\textsubscript{GAN}). For the subject shown in Fig. \ref{impact}, LungViT estimated 32.44$\%$ air trapping and LungViT\textsubscript{EC} estimated 36.88\% air trapping, which were very close to the ground truth value of 34.63$\%$. The LungViT models were able to accurately identify parenchymal regions with air trapping, unlike SwinUNETR\textsubscript{GAN}, which overestimated the spatial extent of the disease with an overall mean of 54.85\% across the entire volume. Fissure segmentations on real and synthetic RV volumes, from pre-trained FissureNet, are shown on sagittal slices in Fig. \ref{fissures}. Differences in quantitative CT biomarkers from real and synthetic RV scans across different GOLD stages has been shown in Fig. \ref{biomarker_and_GOLD}.

We conducted model validation on an out-of-distribution cohort of 200 subjects from the COPDGene study (results shown in Table \ref{copdgene_quant}). The model performed well with PSNR 23.96, SSIM of 0.833 and fSAD error of 4.32\%. A visualization of samples predicted by LungViT\textsubscript{EC} on a subject from COPDGene is shown in Fig. \ref{qual_copdgene}. Lastly, results from the ablation study are presented in Table \ref{ablation}.

\section{Discussion and Conclusion}
Multiple volume surrogates of lung function have gained widespread clinical attention for characterizing local functional abnormalities in COPD~\cite{galban2012computed}. While these measures enable better understanding of disease mechanisms, they require CT images at different volumes which may not be recommended in some clinical settings. We hypothesized that a CT image at end-inspiration contained sufficient structural information to predict the associated aeration change on an expiratory CT image. To that end, we proposed a volumetric texture transfer framework for translating inspiratory CT image intensities to corresponding expiratory CT image intensities.
Although, we demonstrated image translation from a TLC CT to RV, the proposed method can be used to estimate CT intensities at other lung volumes such as the functional residual capacity.

Image-to-image translation for large 3D volumes entailed several challenges. Foremost, was the large GPU memory required to handle increasingly large CT volumes. Most of the existing generative methods address this by training on 2D slices, which leads to discontinuous transitions when slices are stacked together. This limits the radiological utility of synthetic images. We also noticed these discontinuities for 2D GAN models in comparison to their 3D counterparts (see Fig \ref{qual}). To circumvent slicewise discontinuity, we trained our models on 3D patches that were combined at inference by taking the mean along the overlapping boundaries. This helped us estimate a smooth 3D volume by avoiding discontinuities across patches. Patch-based networks often suffer from the limited size of receptive field and may not be able to model global patterns within an image volume. We attempted to address this in several ways. The simplest measure was to use larger patches of size 128 $\times$ 128 $\times$ 128 that were trained on deeper models for increasing the overall receptive field. We modeled the dependencies between distant regions of an image by using a multiresolution Swin Transformer encoder. The encoder was connected at multiple resolutions to a channel attention-based convolutional decoder. We used SE blocks to model relationships between different features learnt by the encoder and the decoder layers. To further alleviate lack of context, we used a coarse-to-fine voxelwise consistency loss and ensemble cascading. The proposed framework, with all its components, was trained using a batch size of 4 on a single NVIDIA A100 GPU with 80 GB memory. The DISTS\textsubscript{MV} module that was proposed to model the stylistic and textural details, entailed a very small proportion of GPU memory and was still able to capture subtle tissue textures enabling perceptually realistic image generation. Unlike the unpaired image-to-image translation methods, which require larger datasets and more GPU memory, we trained our models in a paired setting; which allowed us to develop multiresolution voxel consistency and style-based modules for exploiting one-to-one relationships between paired samples. Image registration before training was first required to ensure point-by-point correspondence between paired TLC and RV images. This increased the reliance of our method on the underlying image registration method; which remains to be investigated in future studies. Although image registration is not required for unpaired approaches such as the CycleGAN, Yang \textit{et al}. recently showed that it improved image synthesis in unpaired settings as well~\cite{yang2020unsupervised}.

Another shortcoming of most volumetric generative models is that they do not attempt to model the inherent style and texture patterns during image translation. Style transfer at a volumetric level is limited due to a lack of 3D models that can be used for generating deep representations. Moreover, stylization typically requires exemplars that are missing for most medical image translation tasks. For our task, modeling texture within the expiratory CT was pertinent since most clinical measures depended largely on variations in local tissue texture. We proposed a multiview texture similarity (DISTS\textsubscript{MV}) to model image textures in 3D. Adding DISTS\textsubscript{MV} significantly improved perceptual image quality when compared to other methods, as shown in Fig. \ref{qual}. Ensembling cascade further improved the overall quantitative performance of the model (as shown in Table \ref{r_1_r2}). 

\begin{figure}[!t]
\centering
\includegraphics[width=0.5\textwidth]{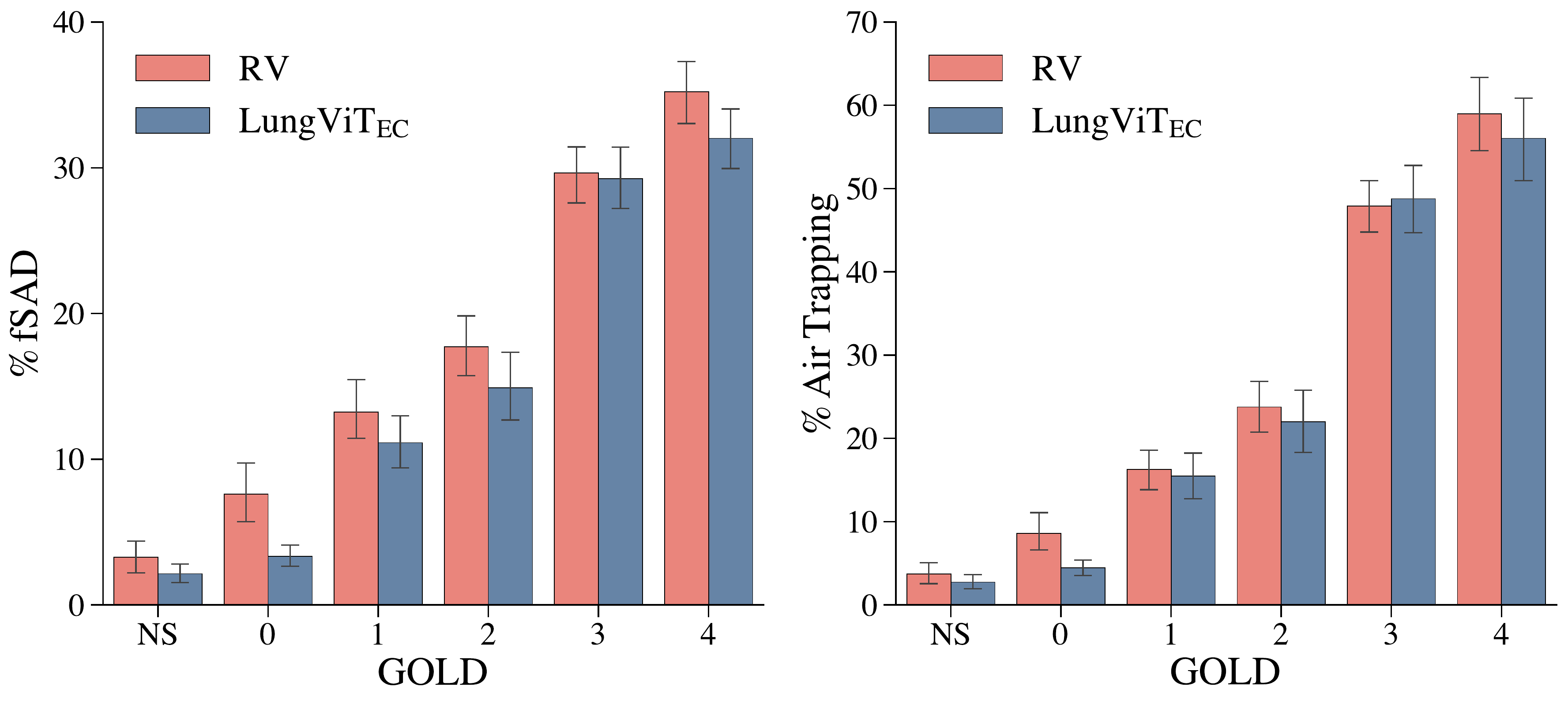}
\vspace{-0.30in}
\caption{Percent fSAD and $\%$ air trapping values across varying degrees of COPD severity, defined by GOLD~\cite{vestbo2013global}. RV denotes $\%$ fSAD and $\%$ air trapping values from real RV scans, deformed to TLC volumes, while LungViT\textsubscript{EC} represents the same biomarkers extracted from synthetic RV volumes generated by the LungViT\textsubscript{EC} framework. `NS' = individuals who never smoked. }
\label{biomarker_and_GOLD}
\end{figure}

\begin{figure}[!t]
\centering
\includegraphics[width=0.48\textwidth]{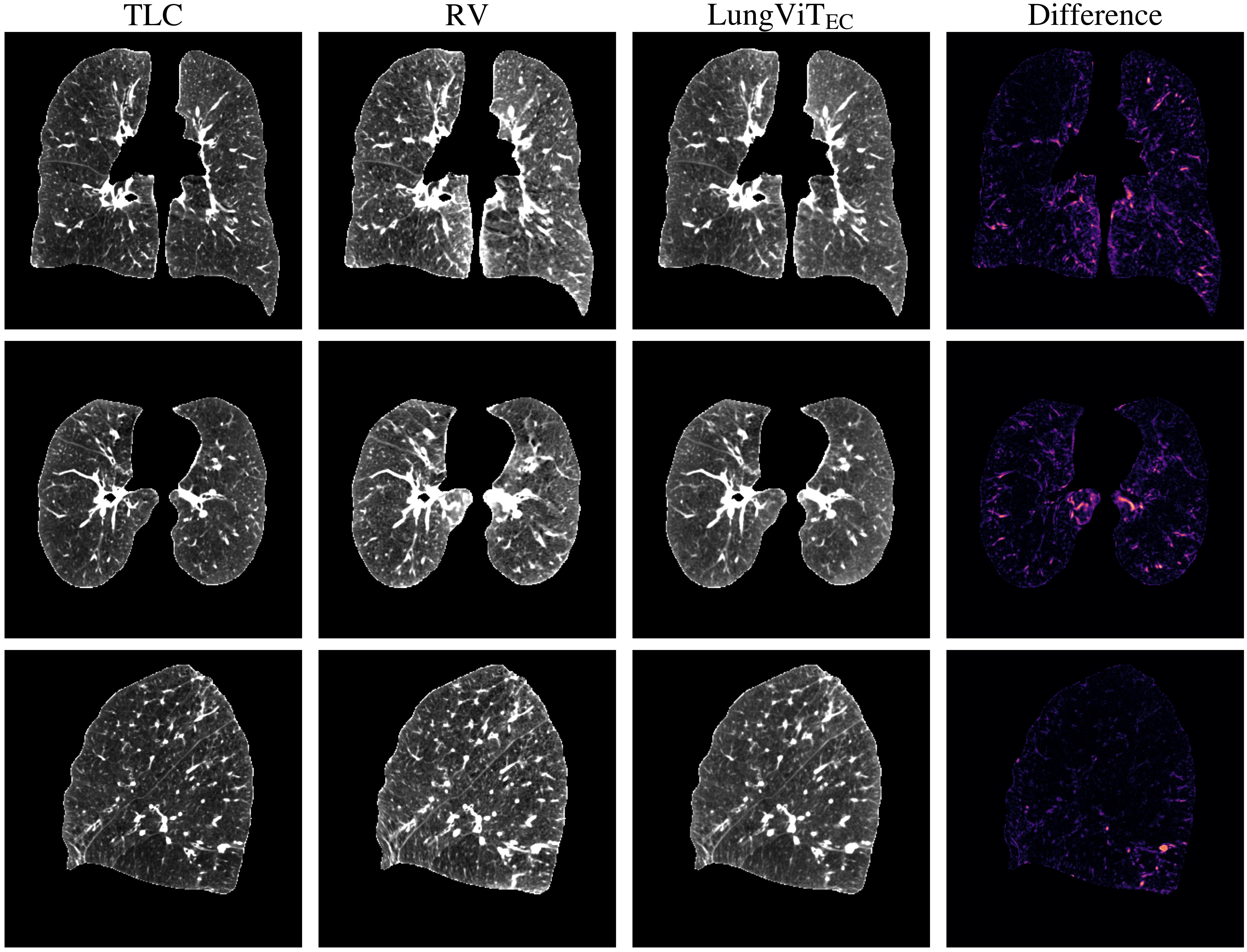}
\vspace{-0.10in}
\caption{Visual performance evaluation of LungViT\textsubscript{EC} on a subject from an out-of-distribution external validation cohort, COPDGene. The difference images represent absolute differences between the ground truth (RV) and predicted LungViT\textsubscript{EC} samples.}
\label{qual_copdgene}
\end{figure}

\begin{table*}[htbp]
    \small
    \centering
    \caption{Ablation study for understand the relative important of various LungViT components. SwinUNETR\textsubscript{GAN} served as the baseline model which was progressively enhanced for the ablation study. We denote the highest metric values in \textbf{black} and second \\ highest values in \textcolor{BlueViolet}{\textbf{blue}}.}
    \vspace{-0.1in}
    \begin{tabularx}{0.95\textwidth}{l l l l l l}
    \toprule
      & \textbf{PSNR (dB)}\textsubscript{$\uparrow$} & \textbf{SSIM}\textsubscript{$\uparrow$} & \textbf{LPIPS}\textsubscript{2.5D}\textsubscript{$\downarrow$} & \textbf{FID}\textsubscript{2.5D}\textsubscript{$\downarrow$}  & \textbf{fSAD}\textsubscript{MAE}\textsubscript{$\downarrow$} \\
    \midrule
    SwinUNETR\textsubscript{GAN} -- baseline & 23.53 $\pm$ 2.11 & 0.823 $\pm$ 0.038 & 0.111 $\pm$ 0.019 & 1.81 $\pm$ 0.47 & 7.55 $\pm$ 7.27 \\
    SwinUNETR\textsubscript{GAN} -- with $\mathcal{L}$\textsubscript{DISTS\textsubscript{MV}} $ + \: \mathcal{L}$\textsubscript{MR} & 24.53 $\pm$ 2.17 & 0.823 $\pm$ 0.038 & 0.099 $\pm$ 0.021 & 1.54 $\pm$ 0.43 & 6.69 $\pm$ 7.54 \\
    LungViT -- with $\mathcal{L}$\textsubscript{DISTS\textsubscript{MV}} $ + \: \mathcal{L}$\textsubscript{MR} $+$ SE blocks & 24.43 $\pm$ 2.42 & 0.837 $\pm$ 0.042 & \textcolor{BlueViolet}{\textbf{0.096}} $\pm$ \textcolor{BlueViolet}{\textbf{0.019}} & \textbf{1.42} $\pm$ \textbf{0.44} & \textbf{6.04} $\pm$ \textbf{7.06} \\
    LungViT with EC  & \textbf{24.75} $\pm$ \textbf{2.47} & \textcolor{BlueViolet}{\textbf{0.842}} $\pm$ \textcolor{BlueViolet}{\textbf{0.040}} & \textbf{0.096} $\pm$ \textbf{0.020} & \textcolor{BlueViolet}{\textbf{1.51}} $\pm$ \textcolor{BlueViolet}{\textbf{0.43}} & \textcolor{BlueViolet}{\textbf{6.06}} $\pm$ \textcolor{BlueViolet}{\textbf{7.07}}  \\
    \bottomrule
\end{tabularx}
    \label{ablation}
\end{table*}

\begin{table}[h]
    \small
    \centering
    \caption{Quantitative performance evaluation of LungViT\textsubscript{EC} on 200 subjects from an out-of-distribution cohort, COPDGene~\cite{regan2011genetic}.}
    \vspace{-0.1in}
    \begin{tabularx}{0.46\textwidth}{l l l l}
    \toprule
      & \textbf{PSNR (dB)}\textsubscript{$\uparrow$} & \textbf{SSIM}\textsubscript{$\uparrow$} & \textbf{fSAD}\textsubscript{MAE}\textsubscript{$\downarrow$} \\
    \midrule
    LungViT\textsubscript{EC} & 23.96 $\pm$ 1.66 & 0.833 $\pm$ 0.028 & 4.32 $\pm$ 4.19  \\
    \bottomrule
\end{tabularx}
    \label{copdgene_quant}
\end{table}

We analyzed the performance of four different state-of-the-art 2D methods in comparison to our work. Two of these models, Pix2Pix~\cite{isola2017image} and the work by Yang \textit{et al}.~\cite{yang2018low}, were based on purely convolutional generators and discriminators; SAGAN was augmented with self-attention blocks and spectral normalization~\cite{zhang2019self}, while PTNet was a pure transformer network~\cite{zhang2022ptnet3d}. Amongst the 2D models, Pix2Pix showed the worst quantitative performance while SAGAN performed the best. Although SAGAN achieved the highest SSIM index of 0.845 (compared to LungViT\textsubscript{EC} with 0.842) shown in Table \ref{r_1_r2}, it showed significant discontinuities across axial and sagittal slices (see Fig. \ref{qual}). It is important to note that PTNet required significantly larger GPU memory, restricting its authors to train it on small patches with a size of 64 $\times$ 64 $\times$ 64~\cite{zhang2022ptnet3d}. While this approach managed memory constraints for their task, it may not have been suitable for synthesizing cross-volume CT images with increasingly large volumes. Therefore, we trained PTNet on 2D coronal slices for comparison where it performed well but also showed slice discontinuities (see Fig. \ref{qual}). We also compared our model with two recent state-of-the-art hybrid generator architectures with transformer and convolutional blocks, UNETR\textsubscript{GAN} and SwinUNETR\textsubscript{GAN} that were trained adversarially. The architecture of our generator, SwinSEER, was a combination of carefully selected convolutional, attentional, and transformer blocks that were collectively able to perform better than most of the recent convolutional and transformer-based generators. The relative importance of each of the framework components was further ascertained by an ablation study shown in Table \ref{ablation}.

We compared the performance of our model to other methods based on two perceptual quality metrics -- LPIPS and FID. The LungViT models showed superior perceptual quality compared to all other models (see LPIPS and FID in Table \ref{r_1_r2}). Although the work of Yang \textit{et al}.~\cite{yang2018low} explicitly minimized LPIPS loss, LungViT was still able to achieve a lower LPIPS value of 0.096 vs. 0.101 (see Table \ref{r_1_r2}). We compared the performance of our models with different methods using a quantitative CT metric-based error -- fSAD\textsubscript{MAE}. Our models achieved lowest errors in fSAD demonstrating highest potential for clinical adaptability.

An important part of our work was constituted by the clinical validation of our model. LungViT\textsubscript{EC} performed well across varying GOLD stages and the model was able to capture changes across more severe disease (see Table \ref{r_2}). Earlier we noted that the 2D models missed subtle changes in tissue textures and produced distorted fissure and vessel structures due to slicewise discontinuities. Unlike these models, intricate vessel structures were captured successfully by LungViT\textsubscript{EC}, as shown in Fig. \ref{q_v_gold}. A Bland Altman analysis between CT biomarkers from real and synthetic scans revealed increased bias of SwinUNETR\textsubscript{GAN} model towards overestimating $\%$ fSAD by 4.84$\%$ and $\%$ air trapping by 5.67$\%$ on average (see top two plots in Fig. \ref{biomarker}). The LungViT model underestimated $\%$ fSAD and $\%$ air trapping with a decreased bias of 3.17$\%$ and 2.29$\%$ respectively (see Fig. \ref{biomarker}). LungViT\textsubscript{EC} underestimated $\%$ fSAD and $\%$ air trapping with minimal bias of just 2.33$\%$ and 1.56$\%$ respectively. Bland Altman analysis showed reliable estimation of clinical biomarkers using the LungViT\textsubscript{EC} framework. A visual comparison in Fig. \ref{impact} also showed improved consistency in spatial distribution of air trapping across real and synthetic samples as compared to the SwinUNETR\textsubscript{GAN}. Identical fissure boundaries identified by an independently trained model showed that our model reliably approximated the real data distribution (see Fig. \ref{fissures}). Model validation through an independently trained network demonstrated realism of the synthetic samples. Clinical validation across different GOLD stages demonstrated that LungViT\textsubscript{EC} was able to predict CT biomarkers (fSAD and air trapping) as disease severity increased (see Fig. \ref{biomarker_and_GOLD}). More importantly, fSAD and air trapping from synthetic RV scans increased with COPD GOLD stages. These results corroborated the clinical findings by Pompe \textit{et al}.~\cite{pompe2020five} and Galban \textit{et al}.~\cite{galban2012computed} that associated COPD GOLD stages with increased air trapping and fSAD respectively. These findings highlighted the potential of synthetically extracted quantitative CT biomarkers for characterizing COPD and related lung disorders. A limitation of our study is that RV shape and volume information is lost since the predicted RV image is anatomically aligned to the TLC image. However, calculation of  CT biomarkers from paired images (e.g. PRM) requires initial image registration, therefore the proposed method can provide these measurements.

Most of the generative modeling studies lack validation on out-of-distribution datasets. We evaluated our model, developed using SPIROMICS, on an external cohort of subjects from COPDGene study. COPDGene acquired CT data using a totally different image acquisition protocol compared to SPIROMICS. Our model showed reliable out-of-distribution generalization with a very small decrease in performance compared to the development cohort (see Table \ref{copdgene_quant}). Interestingly, the fSAD error was lower for COPDGene cohort as compared to SPIROMICS. Qualitative evaluation also showed that model could capture aeration changes from TLC to RV in COPDGene cohort (see Fig. \ref{qual_copdgene}). 

It is common in clinical practice to acquire lung images only at inspiration, and yet it is becoming evident that an early sign of numerous lung diseases is seen on the expiratory image, manifesting as air trapping or fSAD. Our method would allow for the retrospective assessment of patient scans to assess if pathologies were preceded by fSAD, providing critical insights into disease processes. It would enable patient assessment across multiple large cohorts, including MESA~\cite{bild2002multi} and COPDGene~\cite{regan2011genetic}, which acquired CT scans at different volumes. Additionally, the assessment of fSAD is dependent upon a patient making an appropriate effort to expel as much air from the lungs as possible. Furthermore, this difficult maneuver must be repeatable so as to allow for the tracking of disease progression. In many cases, patients with increased symptom burden may not be able to expel much air from the lungs making image acquisition harder and unreliable. LungViT eliminates the need for the patient to achieve a breath hold at this low lung volume, thus allowing for the more accurate tracing of longitudinal changes. Our method also allows for the automated reporting of functional abnormalities across a wide variety of lung diseases providing the potential for expanded insights into early pathologies associated with transplant rejection, prediction of rapid progression of emphysema or fibrosis, and more. In summary, our method allows for retrospective evaluation where multiple lung volumes have not been imaged and prospectively where subject considerations limit the availability of a reliable inspiratory/expiratory image pair.

\vspace{-0.1in}
\section{Compliance with ethical standards}
This research study was conducted retrospectively using anonymized human subject data made publicly available by SPIROMICS. Written consent was provided by all subjects, and the protocols were approved by the Institutional Review Boards (IRBs) of each participating study center.
\vspace{-0.1in}
\section{Acknowledgment}
This work was supported in part by the grant R01HL142625 and by a grant from The Roy J. Carver Charitable Trust. The authors thank the SPIROMICS participants and participating physicians, investigators, and staff for making this research possible. More information about the study and how to access SPIROMICS data is available at www.spiromics.org. The authors would like to acknowledge the University of North Carolina at Chapel Hill BioSpecimen Processing Facility for sample processing, storage, and sample disbursements (http://bsp.web.unc.edu/). We would like to acknowledge the following current and former investigators of the SPIROMICS sites and reading centers: Neil E Alexis, MD; Wayne H Anderson, PhD; Mehrdad Arjomandi, MD; Igor Barjaktarevic, MD, PhD; R Graham Barr, MD, DrPH; Patricia Basta, PhD; Lori A Bateman, MSc; Surya P Bhatt, MD; Eugene R Bleecker, MD; Richard C Boucher, MD; Russell P Bowler, MD, PhD; Stephanie A Christenson, MD; Alejandro P Comellas, MD; Christopher B Cooper, MD, PhD; David J Couper, PhD; Gerard J Criner, MD; Ronald G Crystal, MD; Jeffrey L Curtis, MD; Claire M Doerschuk, MD; Mark T Dransfield, MD; Brad Drummond, MD; Christine M Freeman, PhD; Craig Galban, PhD; MeiLan K Han, MD, MS; Nadia N Hansel, MD, MPH; Annette T Hastie, PhD; Eric A Hoffman, PhD; Yvonne Huang, MD; Robert J Kaner, MD; Richard E Kanner, MD; Eric C Kleerup, MD; Jerry A Krishnan, MD, PhD; Lisa M LaVange, PhD; Stephen C Lazarus, MD; Fernando J Martinez, MD, MS; Deborah A Meyers, PhD; Wendy C Moore, MD; John D Newell Jr, MD; Robert Paine, III, MD; Laura Paulin, MD, MHS; Stephen P Peters, MD, PhD; Cheryl Pirozzi, MD; Nirupama Putcha, MD, MHS; Elizabeth C Oelsner, MD, MPH; Wanda K O’Neal, PhD; Victor E Ortega, MD, PhD; Sanjeev Raman, MBBS, MD; Stephen I. Rennard, MD; Donald P Tashkin, MD; J Michael Wells, MD; Robert A Wise, MD; and Prescott G Woodruff, MD, MPH. The project officers from the Lung Division of the National Heart, Lung, and Blood Institute were Lisa Postow, PhD, and Lisa Viviano, BSN; SPIROMICS was supported by contracts from the NIH/NHL BI (HHSN268200900013C, HHSN268200900014C, HHSN 268200900015C, HHSN268200900016C, HHSN268200900 017C, HHSN268200900018C, HHSN268200900019C, HHS N268200900020C), grants from the NIH/NHLBI (U01 HL1 37880 and U24 HL141762), and supplemented by contributions made through the Foundation for the NIH and the COPD Foundation from AstraZeneca/MedImmune; Bayer; Bellerophon Therapeutics; Boehringer-Ingelheim Pharmaceuticals, Inc.; Chiesi Farmaceutici S.p.A.; Forest Research Institute, Inc.; GlaxoSmithKline; Grifols Therapeutics, Inc.; Ikaria, Inc.; Novartis Pharmaceuticals Corporation; Nycomed GmbH; ProterixBio; Regeneron Pharmaceuticals, Inc.; Sanofi; Sunovion; Takeda Pharmaceutical Company; and Theravance Biopharma and Mylan.

Drs. Hoffman and Reinhardt are shareholders in VIDA Diagnostics, Inc. 
Dr. Christensen has received royalties from VIDA Diagnostics, Inc.
\vspace{-0.1in}
\bibliographystyle{ieeetr}
\bibliography{References}

\begin{thebibliography}{10}

\bibitem{chaudhary2023predicting}
M.~F. Chaudhary, E.~A. Hoffman, J.~Guo, A.~P. Comellas, J.~D. Newell,
  P.~Nagpal, S.~Fortis, G.~E. Christensen, S.~E. Gerard, Y.~Pan, {\em et~al.},
  ``Predicting severe chronic obstructive pulmonary disease exacerbations using
  quantitative ct: a retrospective model development and external validation
  study,'' {\em The Lancet Digital Health}, vol.~5, no.~2, pp.~e83--e92, 2023.

\bibitem{bodduluri2018recent}
S.~Bodduluri, J.~M. Reinhardt, E.~A. Hoffman, J.~D. Newell~Jr, and S.~P. Bhatt,
  ``Recent advances in computed tomography imaging in chronic obstructive
  pulmonary disease,'' {\em Annals of the American Thoracic Society}, vol.~15,
  no.~3, pp.~281--289, 2018.

\bibitem{amudala2023radiomics}
P.~R. Amudala~Puchakayala, V.~L. Sthanam, A.~Nakhmani, M.~F. Chaudhary,
  A.~Kizhakke~Puliyakote, J.~M. Reinhardt, C.~Zhang, S.~P. Bhatt, and
  S.~Bodduluri, ``Radiomics for improved detection of chronic obstructive
  pulmonary disease in low-dose and standard-dose chest {CT} scans,'' {\em
  Radiology}, vol.~307, no.~5, p.~e222998, 2023.

\bibitem{hoffman2022origins}
E.~A. Hoffman, ``Origins of and lessons from quantitative functional {X}-ray
  computed tomography of the lung,'' {\em The British Journal of Radiology},
  vol.~95, no.~1132, p.~20211364, 2022.

\bibitem{reinhardt2008registration}
J.~M. Reinhardt, K.~Ding, K.~Cao, G.~E. Christensen, E.~A. Hoffman, and S.~V.
  Bodas, ``Registration-based estimates of local lung tissue expansion compared
  to xenon {CT} measures of specific ventilation,'' {\em Medical Image
  Analysis}, vol.~12, no.~6, pp.~752--763, 2008.

\bibitem{galban2012computed}
C.~J. Galb{\'a}n, M.~K. Han, J.~L. Boes, K.~A. Chughtai, C.~R. Meyer, T.~D.
  Johnson, S.~Galb{\'a}n, A.~Rehemtulla, E.~A. Kazerooni, F.~J. Martinez, {\em
  et~al.}, ``Computed tomography-based biomarker provides unique signature for
  diagnosis of {COPD} phenotypes and disease progression,'' {\em Nature
  {M}edicine}, vol.~18, no.~11, p.~1711, 2012.

\bibitem{kirby2017novel}
M.~Kirby, Y.~Yin, J.~Tschirren, W.~C. Tan, J.~Leipsic, C.~J. Hague,
  J.~Bourbeau, D.~D. Sin, J.~C. Hogg, H.~O. Coxson, {\em et~al.}, ``A novel
  method of estimating small airway disease using inspiratory-to-expiratory
  computed tomography,'' {\em Respiration}, vol.~94, no.~4, pp.~336--345, 2017.

\bibitem{galban2009parametric}
C.~J. Galb{\'a}n, T.~L. Chenevert, C.~R. Meyer, C.~Tsien, T.~S. Lawrence, D.~A.
  Hamstra, L.~Junck, P.~C. Sundgren, T.~D. Johnson, D.~J. Ross, {\em et~al.},
  ``The parametric response map is an imaging biomarker for early cancer
  treatment outcome,'' {\em Nature Medicine}, vol.~15, no.~5, pp.~572--576,
  2009.

\bibitem{bodduluri2017biomechanical}
S.~Bodduluri, S.~P. Bhatt, E.~A. Hoffman, J.~D. Newell, C.~H. Martinez, M.~T.
  Dransfield, M.~K. Han, and J.~M. Reinhardt, ``Biomechanical {CT} metrics are
  associated with patient outcomes in {COPD},'' {\em Thorax}, vol.~72, no.~5,
  pp.~409--414, 2017.

\bibitem{trivedi2022quantitative}
A.~P. Trivedi, C.~Hall, C.~W. Goss, D.~Lew, J.~G. Krings, M.~C. McGregor,
  M.~Samant, J.~P. Sieren, H.~Li, K.~B. Schechtman, {\em et~al.},
  ``Quantitative {CT} characteristics of cluster phenotypes in the severe
  asthma research program cohorts,'' {\em Radiology}, vol.~304, no.~2,
  pp.~450--459, 2022.

\bibitem{sieren2016spiromics}
J.~P. Sieren, J.~D. Newell~Jr, R.~G. Barr, E.~R. Bleecker, N.~Burnette, E.~E.
  Carretta, D.~Couper, J.~Goldin, J.~Guo, M.~K. Han, {\em et~al.},
  ``{SPIROMICS} protocol for multicenter quantitative computed tomography to
  phenotype the lungs,'' {\em American Journal of Respiratory and Critical Care
  Medicine}, vol.~194, no.~7, pp.~794--806, 2016.

\bibitem{newell2013development}
J.~D. Newell~Jr, J.~Sieren, and E.~A. Hoffman, ``Development of quantitative ct
  lung protocols,'' {\em Journal of thoracic imaging}, vol.~28, no.~5, 2013.

\bibitem{bild2002multi}
D.~E. Bild, D.~A. Bluemke, G.~L. Burke, R.~Detrano, A.~V. Diez~Roux, A.~R.
  Folsom, P.~Greenland, D.~R. JacobsJr, R.~Kronmal, K.~Liu, {\em et~al.},
  ``Multi-ethnic study of atherosclerosis: objectives and design,'' {\em
  American journal of epidemiology}, vol.~156, no.~9, pp.~871--881, 2002.

\bibitem{regan2011genetic}
E.~A. Regan, J.~E. Hokanson, J.~R. Murphy, B.~Make, D.~A. Lynch, T.~H. Beaty,
  D.~Curran-Everett, E.~K. Silverman, and J.~D. Crapo, ``Genetic epidemiology
  of {COPD} ({COPDG}ene) study design,'' {\em COPD: Journal of Chronic
  Obstructive Pulmonary Disease}, vol.~7, no.~1, pp.~32--43, 2011.

\bibitem{wolterink2017deep}
J.~M. Wolterink, A.~M. Dinkla, M.~H. Savenije, P.~R. Seevinck, C.~A. van~den
  Berg, and I.~I{\v{s}}gum, ``Deep {MR} to {CT} synthesis using unpaired
  data,'' in {\em International Workshop on Simulation and Synthesis in Medical
  Imaging}, pp.~14--23, Springer, 2017.

\bibitem{nie2018medical}
D.~Nie, R.~Trullo, J.~Lian, L.~Wang, C.~Petitjean, S.~Ruan, Q.~Wang, and
  D.~Shen, ``Medical image synthesis with deep convolutional adversarial
  networks,'' {\em IEEE Transactions on Biomedical Engineering}, vol.~65,
  no.~12, pp.~2720--2730, 2018.

\bibitem{yu2019ea}
B.~Yu, L.~Zhou, L.~Wang, Y.~Shi, J.~Fripp, and P.~Bourgeat, ``Ea-{GAN}s:
  Edge-aware generative adversarial networks for cross-modality {MR} image
  synthesis,'' {\em IEEE Transactions on Medical Imaging}, vol.~38, no.~7,
  pp.~1750--1762, 2019.

\bibitem{yang2020unsupervised}
H.~Yang, J.~Sun, A.~Carass, C.~Zhao, J.~Lee, J.~L. Prince, and Z.~Xu,
  ``Unsupervised {MR}-to-{CT} synthesis using structure-constrained
  {C}ycle{GAN},'' {\em IEEE Transactions on Medical Imaging}, vol.~39, no.~12,
  pp.~4249--4261, 2020.

\bibitem{yu2020sample}
B.~Yu, L.~Zhou, L.~Wang, Y.~Shi, J.~Fripp, and P.~Bourgeat, ``Sample-adaptive
  {GAN}s: linking global and local mappings for cross-modality {MR} image
  synthesis,'' {\em IEEE Transactions on Medical Imaging}, vol.~39, no.~7,
  pp.~2339--2350, 2020.

\bibitem{zhou2020hi}
T.~Zhou, H.~Fu, G.~Chen, J.~Shen, and L.~Shao, ``Hi-{N}et: Hybrid-fusion
  network for multi-modal {MR} image synthesis,'' {\em IEEE Transactions on
  Medical Imaging}, vol.~39, no.~9, pp.~2772--2781, 2020.

\bibitem{dar2019image}
S.~U. Dar, M.~Yurt, L.~Karacan, A.~Erdem, E.~Erdem, and T.~{\c{C}}ukur, ``Image
  synthesis in multi-contrast {MRI} with conditional generative adversarial
  networks,'' {\em IEEE Transactions on Medical Imaging}, vol.~38, no.~10,
  pp.~2375--2388, 2019.

\bibitem{emami2018generating}
H.~Emami, M.~Dong, S.~P. Nejad-Davarani, and C.~K. Glide-Hurst, ``Generating
  synthetic {CT}s from magnetic resonance images using generative adversarial
  networks,'' {\em Medical Physics}, vol.~45, no.~8, pp.~3627--3636, 2018.

\bibitem{uzunova2020memory}
H.~Uzunova, J.~Ehrhardt, and H.~Handels, ``Memory-efficient {GAN}-based domain
  translation of high resolution 3{D} medical images,'' {\em Computerized
  Medical Imaging and Graphics}, vol.~86, p.~101801, 2020.

\bibitem{luo20213d}
Y.~Luo, Y.~Wang, C.~Zu, B.~Zhan, X.~Wu, J.~Zhou, D.~Shen, and L.~Zhou, ``3d
  transformer-gan for high-quality pet reconstruction,'' in {\em Medical Image
  Computing and Computer Assisted Intervention--MICCAI 2021: 24th International
  Conference, Strasbourg, France, September 27--October 1, 2021, Proceedings,
  Part VI 24}, pp.~276--285, Springer, 2021.

\bibitem{zhang2021transct}
Z.~Zhang, L.~Yu, X.~Liang, W.~Zhao, and L.~Xing, ``Transct: dual-path
  transformer for low dose computed tomography,'' in {\em Medical Image
  Computing and Computer Assisted Intervention--MICCAI 2021: 24th International
  Conference, Strasbourg, France, September 27--October 1, 2021, Proceedings,
  Part VI 24}, pp.~55--64, Springer, 2021.

\bibitem{korkmaz2022unsupervised}
Y.~Korkmaz, S.~U. Dar, M.~Yurt, M.~{\"O}zbey, and T.~Cukur, ``Unsupervised mri
  reconstruction via zero-shot learned adversarial transformers,'' {\em IEEE
  Transactions on Medical Imaging}, vol.~41, no.~7, pp.~1747--1763, 2022.

\bibitem{you2019ct}
C.~You, G.~Li, Y.~Zhang, X.~Zhang, H.~Shan, M.~Li, S.~Ju, Z.~Zhao, Z.~Zhang,
  W.~Cong, {\em et~al.}, ``{CT} super-resolution {GAN} constrained by the
  identical, residual, and cycle learning ensemble ({GAN-CIRCLE}),'' {\em IEEE
  Transactions on Medical Imaging}, vol.~39, no.~1, pp.~188--203, 2019.

\bibitem{you2022fine}
S.~You, B.~Lei, S.~Wang, C.~K. Chui, A.~C. Cheung, Y.~Liu, M.~Gan, G.~Wu, and
  Y.~Shen, ``Fine perceptive {GAN}s for brain {MR} image super-resolution in
  wavelet domain,'' {\em IEEE Transactions on Neural Networks and Learning
  Systems}, 2022.

\bibitem{lyu2020mri}
Q.~Lyu, H.~Shan, and G.~Wang, ``{MRI} super-resolution with ensemble learning
  and complementary priors,'' {\em IEEE Transactions on Computational Imaging},
  vol.~6, pp.~615--624, 2020.

\bibitem{yang2018low}
Q.~Yang, P.~Yan, Y.~Zhang, H.~Yu, Y.~Shi, X.~Mou, M.~K. Kalra, Y.~Zhang,
  L.~Sun, and G.~Wang, ``Low-dose {CT} image denoising using a generative
  adversarial network with {W}asserstein distance and perceptual loss,'' {\em
  IEEE Transactions on Medical Imaging}, vol.~37, no.~6, pp.~1348--1357, 2018.

\bibitem{bera2021noise}
S.~Bera and P.~K. Biswas, ``Noise conscious training of non local neural
  network powered by self attentive spectral normalized markovian patch {GAN}
  for low dose {CT} denoising,'' {\em IEEE Transactions on Medical Imaging},
  vol.~40, no.~12, pp.~3663--3673, 2021.

\bibitem{shin2018medical}
H.~C. Shin, N.~A. Tenenholtz, J.~K. Rogers, C.~G. Schwarz, M.~L. Senjem, J.~L.
  Gunter, K.~P. Andriole, and M.~Michalski, ``Medical image synthesis for data
  augmentation and anonymization using generative adversarial networks,'' in
  {\em International Workshop on Simulation and Synthesis in Medical Imaging},
  pp.~1--11, Springer, 2018.

\bibitem{frid2018gan}
M.~Frid-Adar, I.~Diamant, E.~Klang, M.~Amitai, J.~Goldberger, and H.~Greenspan,
  ``{GAN}-based synthetic medical image augmentation for increased {CNN}
  performance in liver lesion classification,'' {\em Neurocomputing}, vol.~321,
  pp.~321--331, 2018.

\bibitem{kustner2019retrospective}
T.~K{\"u}stner, K.~Armanious, J.~Yang, B.~Yang, F.~Schick, and S.~Gatidis,
  ``Retrospective correction of motion-affected mr images using deep learning
  frameworks,'' {\em Magnetic resonance in medicine}, vol.~82, no.~4,
  pp.~1527--1540, 2019.

\bibitem{nie2017medical}
D.~Nie, R.~Trullo, J.~Lian, C.~Petitjean, S.~Ruan, Q.~Wang, and D.~Shen,
  ``Medical image synthesis with context-aware generative adversarial
  networks,'' in {\em International Conference on Medical Image Computing and
  Computer-Assisted Intervention}, pp.~417--425, Springer, 2017.

\bibitem{dosovitskiy2020image}
A.~Dosovitskiy, L.~Beyer, A.~Kolesnikov, D.~Weissenborn, X.~Zhai,
  T.~Unterthiner, M.~Dehghani, M.~Minderer, G.~Heigold, S.~Gelly, {\em et~al.},
  ``An image is worth 16x16 words: Transformers for image recognition at
  scale,'' in {\em International Conference on Learning Representations}, 2020.

\bibitem{kamran2021vtgan}
S.~A. Kamran, K.~F. Hossain, A.~Tavakkoli, S.~L. Zuckerbrod, and S.~A. Baker,
  ``Vtgan: Semi-supervised retinal image synthesis and disease prediction using
  vision transformers,'' in {\em Proceedings of the IEEE/CVF International
  Conference on Computer Vision}, pp.~3235--3245, 2021.

\bibitem{dalmaz2022resvit}
O.~Dalmaz, M.~Yurt, and T.~{\c{C}}ukur, ``Resvit: Residual vision transformers
  for multimodal medical image synthesis,'' {\em IEEE Transactions on Medical
  Imaging}, vol.~41, no.~10, pp.~2598--2614, 2022.

\bibitem{zhang2022ptnet3d}
X.~Zhang, X.~He, J.~Guo, N.~Ettehadi, N.~Aw, D.~Semanek, J.~Posner, A.~Laine,
  and Y.~Wang, ``Ptnet3d: A 3d high-resolution longitudinal infant brain mri
  synthesizer based on transformers,'' {\em IEEE Transactions on Medical
  Imaging}, vol.~41, no.~10, pp.~2925--2940, 2022.

\bibitem{ding2020image}
K.~Ding, K.~Ma, S.~Wang, and E.~P. Simoncelli, ``Image quality assessment:
  Unifying structure and texture similarity,'' {\em IEEE Transactions on
  Pattern Analysis and Machine Intelligence}, vol.~44, no.~5, pp.~2567--2581,
  2020.

\bibitem{gerard2020multi}
S.~E. Gerard, J.~Herrmann, D.~W. Kaczka, G.~Musch, A.~Fernandez-Bustamante, and
  J.~M. Reinhardt, ``Multi-resolution convolutional neural networks for fully
  automated segmentation of acutely injured lungs in multiple species,'' {\em
  Medical Image Analysis}, vol.~60, p.~101592, 2020.

\bibitem{armanious2020medgan}
K.~Armanious, C.~Jiang, M.~Fischer, T.~K{\"u}stner, T.~Hepp, K.~Nikolaou,
  S.~Gatidis, and B.~Yang, ``Med{GAN}: Medical image translation using
  {GAN}s,'' {\em Computerized Medical Imaging and Graphics}, vol.~79,
  p.~101684, 2020.

\bibitem{kustner2021deep}
T.~K{\"u}stner, C.~Munoz, A.~Psenicny, A.~Bustin, N.~Fuin, H.~Qi, R.~Neji,
  K.~Kunze, R.~Hajhosseiny, C.~Prieto, {\em et~al.}, ``Deep-learning based
  super-resolution for 3d isotropic coronary mr angiography in less than a
  minute,'' {\em Magnetic Resonance in Medicine}, vol.~86, no.~5,
  pp.~2837--2852, 2021.

\bibitem{chaudhary2022single}
M.~F. Chaudhary, S.~E. Gerard, D.~Wang, G.~E. Christensen, C.~B. Cooper, J.~D.
  Schroeder, E.~A. Hoffman, and J.~M. Reinhardt, ``Single volume lung
  biomechanics from chest computed tomography using a mode preserving
  generative adversarial network,'' in {\em 2022 IEEE 19th International
  Symposium on Biomedical Imaging (ISBI)}, pp.~1--5, IEEE, 2022.

\bibitem{zhu2017unpaired}
J.-Y. Zhu, T.~Park, P.~Isola, and A.~A. Efros, ``Unpaired image-to-image
  translation using cycle-consistent adversarial networks,'' in {\em
  Proceedings of the IEEE International Conference on Computer Vision},
  pp.~2223--2232, 2017.

\bibitem{tu2009auto}
Z.~Tu and X.~Bai, ``Auto-context and its application to high-level vision tasks
  and 3{D} brain image segmentation,'' {\em IEEE Transactions on Pattern
  Analysis and Machine Intelligence}, vol.~32, no.~10, pp.~1744--1757, 2009.

\bibitem{chen2020mri}
Y.~Chen, A.~G. Christodoulou, Z.~Zhou, F.~Shi, Y.~Xie, and D.~Li, ``Mri
  super-resolution with gan and 3d multi-level densenet: smaller, faster, and
  better,'' {\em arXiv preprint arXiv:2003.01217}, 2020.

\bibitem{pham2019multiscale}
C.-H. Pham, C.~Tor-D{\'\i}ez, H.~Meunier, N.~Bednarek, R.~Fablet, N.~Passat,
  and F.~Rousseau, ``Multiscale brain mri super-resolution using deep 3d
  convolutional networks,'' {\em Computerized Medical Imaging and Graphics},
  vol.~77, p.~101647, 2019.

\bibitem{isola2017image}
P.~Isola, J.-Y. Zhu, T.~Zhou, and A.~A. Efros, ``Image-to-image translation
  with conditional adversarial networks,'' in {\em Proceedings of the IEEE
  Conference on Computer Vision and Pattern Recognition}, pp.~1125--1134, 2017.

\bibitem{johnson2019conditional}
P.~M. Johnson and M.~Drangova, ``Conditional generative adversarial network for
  3d rigid-body motion correction in mri,'' {\em Magnetic Resonance in
  Medicine}, vol.~82, no.~3, pp.~901--910, 2019.

\bibitem{gerard2023direct}
S.~E. Gerard, M.~F. Chaudhary, J.~Herrmann, G.~E. Christensen, R.~San
  Jose~Estepar, J.~M. Reinhardt, and E.~A. Hoffman, ``Direct estimation of
  regional lung volume change from paired and single ct images using residual
  regression neural network,'' {\em Medical Physics}, 2023.

\bibitem{zhang2019self}
H.~Zhang, I.~Goodfellow, D.~Metaxas, and A.~Odena, ``Self-attention generative
  adversarial networks,'' in {\em International conference on machine
  learning}, pp.~7354--7363, PMLR, 2019.

\bibitem{li2020sacnn}
M.~Li, W.~Hsu, X.~Xie, J.~Cong, and W.~Gao, ``Sacnn: Self-attention
  convolutional neural network for low-dose ct denoising with self-supervised
  perceptual loss network,'' {\em IEEE Transactions on Medical Imaging},
  vol.~39, no.~7, pp.~2289--2301, 2020.

\bibitem{chen2021transunet}
J.~Chen, Y.~Lu, Q.~Yu, X.~Luo, E.~Adeli, Y.~Wang, L.~Lu, A.~L. Yuille, and
  Y.~Zhou, ``Transunet: Transformers make strong encoders for medical image
  segmentation,'' {\em arXiv preprint arXiv:2102.04306}, 2021.

\bibitem{xie2021cotr}
Y.~Xie, J.~Zhang, C.~Shen, and Y.~Xia, ``Cotr: Efficiently bridging cnn and
  transformer for 3d medical image segmentation,'' in {\em Medical Image
  Computing and Computer Assisted Intervention--MICCAI 2021: 24th International
  Conference, Strasbourg, France, September 27--October 1, 2021, Proceedings,
  Part III 24}, pp.~171--180, Springer, 2021.

\bibitem{hatamizadeh2022unetr}
A.~Hatamizadeh, Y.~Tang, V.~Nath, D.~Yang, A.~Myronenko, B.~Landman, H.~R.
  Roth, and D.~Xu, ``Unetr: Transformers for 3d medical image segmentation,''
  in {\em Proceedings of the IEEE/CVF winter Conference on Applications of
  Computer Vision}, pp.~574--584, 2022.

\bibitem{shin2020ganbert}
H.-C. Shin, A.~Ihsani, S.~Mandava, S.~T. Sreenivas, C.~Forster, J.~Cha, and
  A.~D.~N. Initiative, ``Ganbert: Generative adversarial networks with
  bidirectional encoder representations from transformers for mri to pet
  synthesis,'' {\em arXiv preprint arXiv:2008.04393}, 2020.

\bibitem{li2023multi}
Y.~Li, T.~Zhou, K.~He, Y.~Zhou, and D.~Shen, ``Multi-scale transformer network
  with edge-aware pre-training for cross-modality mr image synthesis,'' {\em
  IEEE Transactions on Medical Imaging}, 2023.

\bibitem{chen2022transmorph}
J.~Chen, E.~C. Frey, Y.~He, W.~P. Segars, Y.~Li, and Y.~Du, ``Transmorph:
  Transformer for unsupervised medical image registration,'' {\em Medical Image
  Analysis}, vol.~82, p.~102615, 2022.

\bibitem{shi2022xmorpher}
J.~Shi, Y.~He, Y.~Kong, J.-L. Coatrieux, H.~Shu, G.~Yang, and S.~Li,
  ``Xmorpher: Full transformer for deformable medical image registration via
  cross attention,'' in {\em International Conference on Medical Image
  Computing and Computer-Assisted Intervention}, pp.~217--226, Springer, 2022.

\bibitem{zhang2018unreasonable}
R.~Zhang, P.~Isola, A.~A. Efros, E.~Shechtman, and O.~Wang, ``The unreasonable
  effectiveness of deep features as a perceptual metric,'' in {\em Proceedings
  of the IEEE Conference on Computer Vision and Pattern Recognition},
  pp.~586--595, 2018.

\bibitem{liu2021style}
M.~Liu, P.~Maiti, S.~Thomopoulos, A.~Zhu, Y.~Chai, H.~Kim, and N.~Jahanshad,
  ``Style transfer using generative adversarial networks for multi-site mri
  harmonization,'' in {\em Medical Image Computing and Computer Assisted
  Intervention--MICCAI 2021: 24th International Conference, Strasbourg, France,
  September 27--October 1, 2021, Proceedings, Part III 24}, pp.~313--322,
  Springer, 2021.

\bibitem{goodfellow2014generative}
I.~Goodfellow, J.~Pouget-Abadie, M.~Mirza, B.~Xu, D.~Warde-Farley, S.~Ozair,
  A.~Courville, and Y.~Bengio, ``Generative adversarial networks,'' in {\em
  Advances in Neural Information Processing Systems}, pp.~2672--2680, 2014.

\bibitem{yin2009mass}
Y.~Yin, E.~A. Hoffman, and C.-L. Lin, ``Mass preserving nonrigid registration
  of ct lung images using cubic b-spline,'' {\em Medical physics}, vol.~36,
  no.~9Part1, pp.~4213--4222, 2009.

\bibitem{mao2017least}
X.~Mao, Q.~Li, H.~Xie, R.~Y. Lau, Z.~Wang, and S.~Paul~Smolley, ``Least squares
  generative adversarial networks,'' in {\em Proceedings of the IEEE
  international conference on computer vision}, pp.~2794--2802, 2017.

\bibitem{liu2021swin}
Z.~Liu, Y.~Lin, Y.~Cao, H.~Hu, Y.~Wei, Z.~Zhang, S.~Lin, and B.~Guo, ``Swin
  transformer: Hierarchical vision transformer using shifted windows,'' in {\em
  Proceedings of the IEEE/CVF International Conference on Computer Vision},
  pp.~10012--10022, 2021.

\bibitem{hatamizadeh2021swin}
A.~Hatamizadeh, V.~Nath, Y.~Tang, D.~Yang, H.~R. Roth, and D.~Xu, ``Swin unetr:
  Swin transformers for semantic segmentation of brain tumors in mri images,''
  in {\em International MICCAI Brainlesion Workshop}, pp.~272--284, Springer,
  2021.

\bibitem{hu2018squeeze}
J.~Hu, L.~Shen, and G.~Sun, ``Squeeze-and-excitation networks,'' in {\em
  Proceedings of the IEEE Conference on Computer Vision and Pattern
  Recognition}, pp.~7132--7141, 2018.

\bibitem{wang2004image}
Z.~Wang, A.~C. Bovik, H.~R. Sheikh, and E.~P. Simoncelli, ``Image quality
  assessment: from error visibility to structural similarity,'' {\em IEEE
  Transactions on Image Processing}, vol.~13, no.~4, pp.~600--612, 2004.

\bibitem{deng2009imagenet}
J.~Deng, W.~Dong, R.~Socher, L.-J. Li, K.~Li, and L.~Fei-Fei, ``Imagenet: A
  large-scale hierarchical image database,'' in {\em 2009 IEEE conference on
  computer vision and pattern recognition}, pp.~248--255, Ieee, 2009.

\bibitem{couper2014design}
D.~Couper, L.~M. LaVange, M.~Han, R.~G. Barr, E.~Bleecker, E.~A. Hoffman,
  R.~Kanner, E.~Kleerup, F.~J. Martinez, P.~G. Woodruff, {\em et~al.}, ``Design
  of the {S}ub{P}opulations and {I}ntermediate {O}utcomes in {COPD} {S}tudy
  ({SPIROMICS}),'' {\em Thorax}, vol.~69, no.~5, pp.~492--495, 2014.

\bibitem{vestbo2013global}
J.~Vestbo, S.~S. Hurd, A.~G. Agust{\'\i}, P.~W. Jones, C.~Vogelmeier,
  A.~Anzueto, P.~J. Barnes, L.~M. Fabbri, F.~J. Martinez, M.~Nishimura, {\em
  et~al.}, ``Global strategy for the diagnosis, management, and prevention of
  chronic obstructive pulmonary disease: {GOLD} executive summary,'' {\em
  American Journal of Respiratory and Critical Care Medicine}, vol.~187, no.~4,
  pp.~347--365, 2013.

\bibitem{gorbunova2012mass}
V.~Gorbunova, J.~Sporring, P.~Lo, M.~Loeve, H.~A. Tiddens, M.~Nielsen,
  A.~Dirksen, and M.~de~Bruijne, ``Mass preserving image registration for lung
  {CT},'' {\em Medical Image Analysis}, vol.~16, no.~4, pp.~786--795, 2012.

\bibitem{cao2010regularized}
K.~Cao, K.~Du, K.~Ding, J.~M. Reinhardt, and G.~E. Christensen, ``Regularized
  nonrigid registration of lung {CT} images by preserving tissue volume and
  vesselness measure,'' {\em Grand Challenges in Medical Image Analysis},
  pp.~43--54, 2010.

\bibitem{heusel2017gans}
M.~Heusel, H.~Ramsauer, T.~Unterthiner, B.~Nessler, and S.~Hochreiter, ``{GAN}s
  trained by a two time-scale update rule converge to a local {N}ash
  equilibrium,'' in {\em Advances in Neural Information Processing Systems},
  pp.~6626--6637, 2017.

\bibitem{giavarina2015understanding}
D.~Giavarina, ``Understanding {B}land {A}ltman analysis,'' {\em Biochemia
  Medica}, vol.~25, no.~2, pp.~141--151, 2015.

\bibitem{gerard2018fissurenet}
S.~E. Gerard, T.~J. Patton, G.~E. Christensen, J.~E. Bayouth, and J.~M.
  Reinhardt, ``Fissure{N}et: A deep learning approach for pulmonary fissure
  detection in {CT} images,'' {\em IEEE {T}ransactions on {M}edical {I}maging},
  vol.~38, no.~1, pp.~156--166, 2018.

\bibitem{dice1945measures}
L.~R. Dice, ``Measures of the amount of ecologic association between species,''
  {\em Ecology}, vol.~26, no.~3, pp.~297--302, 1945.

\bibitem{jaccard1912distribution}
P.~Jaccard, ``The distribution of the flora in the alpine zone. 1,'' {\em New
  phytologist}, vol.~11, no.~2, pp.~37--50, 1912.

\bibitem{heimann2009comparison}
T.~Heimann, B.~Van~Ginneken, M.~A. Styner, Y.~Arzhaeva, V.~Aurich, C.~Bauer,
  A.~Beck, C.~Becker, R.~Beichel, G.~Bekes, {\em et~al.}, ``Comparison and
  evaluation of methods for liver segmentation from {CT} datasets,'' {\em IEEE
  Transactions on Medical Imaging}, vol.~28, no.~8, pp.~1251--1265, 2009.

\bibitem{paszke2019pytorch}
A.~Paszke, S.~Gross, F.~Massa, A.~Lerer, J.~Bradbury, G.~Chanan, T.~Killeen,
  Z.~Lin, N.~Gimelshein, L.~Antiga, {\em et~al.}, ``Py{T}orch: An imperative
  style, high-performance deep learning library,'' {\em Advances in Neural
  Information Processing Systems}, vol.~32, 2019.

\bibitem{cardoso2022monai}
M.~J. Cardoso, W.~Li, R.~Brown, N.~Ma, E.~Kerfoot, Y.~Wang, B.~Murrey,
  A.~Myronenko, C.~Zhao, D.~Yang, {\em et~al.}, ``{MONAI}: An open-source
  framework for deep learning in healthcare,'' {\em arXiv preprint
  arXiv:2211.02701}, 2022.

\bibitem{newman1994quantitative}
K.~B. Newman, D.~A. Lynch, L.~S. Newman, D.~Ellegood, and J.~D. Newell~Jr,
  ``Quantitative computed tomography detects air trapping due to asthma,'' {\em
  Chest}, vol.~106, no.~1, pp.~105--109, 1994.

\bibitem{pompe2020five}
E.~Pompe, M.~Strand, E.~M. van Rikxoort, E.~A. Hoffman, R.~G. Barr, J.~P.
  Charbonnier, S.~Humphries, M.~K. Han, J.~E. Hokanson, B.~J. Make, {\em
  et~al.}, ``Five-year progression of emphysema and air trapping at {CT} in
  smokers with and those without chronic obstructive pulmonary disease: results
  from the {COPDG}ene study,'' {\em Radiology}, vol.~295, no.~1, p.~218, 2020.

\end{thebibliography}

\end{document}